\documentclass[%
 preprint,
 amsmath,amssymb,
 aps, physrev,showkeys,floatfix
]{revtex4-2}
\usepackage{caption}
\usepackage{placeins}
\usepackage{graphicx}
\usepackage{dcolumn}
\usepackage{bm}
\usepackage{amsmath}
\usepackage{url}
\usepackage{color}
\usepackage{placeins}
\usepackage{siunitx}
\usepackage[utf8]{inputenc}
\usepackage{comment}
\usepackage{mathtools}
\usepackage{braket}
\usepackage{subcaption}
\usepackage{svg}
\usepackage{xcolor}
\usepackage{float}
\usepackage{multirow}
\usepackage[dvipsnames]{xcolor}
\usepackage{hyperref}
\usepackage[nameinlink]{cleveref}
\usepackage{soul}
\usepackage[table]{xcolor}
\Crefname{figure}{Fig.}{Figs.}

\usepackage{subcaption}
\usepackage{tikz}
\usepackage{pgfplots}
\usepackage{pgfplotstable}
\usepgfplotslibrary{groupplots}
\pgfplotsset{compat=1.18}

\begin{document}
\usetikzlibrary{calc,positioning,arrows.meta}

\preprint{APS/123-QED}

\title{\textbf{Predictive beam-lattice reduction for higher-order topological modes in a 2D SSH phononic crystal}}%

\author{Eloi Pérez Compte}
\affiliation{%
 Department of Applied Science and Technology, Politecnico di Torino, \\Corso Duca degli Abruzzi 24, 10129
Torino, Italy
}%

\author{Michele Brun}
\affiliation{
Department of Mechanical, Chemical and Materials Engineering, University of Cagliari, \\Piazza d’Armi, 09123, Cagliari, Italy
}

\author{Giorgio Carta}
\affiliation{
Department of Mechanical, Chemical and Materials Engineering, University of Cagliari, \\Piazza d’Armi, 09123, Cagliari, Italy
}

\author{Nicola M. Pugno}
\affiliation{
Mechano-X Labs, Department of Civil, Environmental and Mechanical Engineering, University of Trento,, \\Via Mesiano, 77, Trento 38123, Italy
}

\author{Antonio S. Gliozzi}
\affiliation{
 Department of Applied Science and Technology, Politecnico di Torino, \\Corso Duca degli Abruzzi 24, 10129
Torino, Italy
}%

\author{Federico Bosia}
\email{Contact author: federico.bosia@polito.it}
\affiliation{
 Department of Applied Science and Technology, Politecnico di Torino, \\Corso Duca degli Abruzzi 24, 10129
Torino, Italy
}%


\date{\today}

\begin{abstract}
\noindent

\noindent
We develop a mechanically faithful reduced model for a two-dimensional topological phononic crystal composed of rigid square masses connected by slender elastic ligaments. Exploiting Euler–Bernoulli beam theory, we derive a Hermitian 12-degree-of-freedom dynamical matrix that retains in-plane translations, rotations, and ligament eccentricity. This reduction captures effects that are absent from scalar mass–spring SSH models while remaining computationally much more tractable and more easily interpretable than full finite-element simulations. Dimerizing the ligament widths produces a mechanical 2D SSH lattice with a full band gap and a quantized bulk polarization. The sign of the dimerization controls the transition from trivial to non-trivial phases, while ligament eccentricity provides an additional purely geometric mechanism for changing the topology. Ribbon and finite-cell calculations predict in-gap edge and corner modes, quantified by localization measures and confirmed by finite-element simulations. Measurements on 3D-printed samples show an evanescent response in the trivial structure and enhanced boundary/corner response in the non-trivial structure within the predicted gap. The results provide a validated route for designing topological elastic metamaterials using a continuum-informed discrete model rather than either idealized mass–spring networks or brute-force numerical optimization.   

\end{abstract}

\maketitle


\section{Introduction}\label{sec:introduction}


In condensed matter physics, topological insulators (TI) are a class of materials that exhibit a spectral gap in which topologically protected modes can occur \cite{qi2011topological}. This band gap can be labelled as topologically trivial or non-trivial if  the associated topological invariant is zero or non-zero, respectively \cite{shen2012topological}. Such materials obey bulk-edge correspondence, whereby having a bulk band structure with a non-trivial band gap enables localized edge states \cite{qi2006general}. The edge states are protected by the topology, which means that they are robust to defects or disorder \cite{harari2018topological}. Their first realization can be traced to the quantum Hall effect, where an insulating bulk supports chiral edge channels with suppressed backscattering. The Hall conductivity is quantized as \(\sigma_{xy}=C e^2/h\), where the integer \(C\) is the Chern number and constitutes the topological invariant of the system \cite{von1986quantized, laughlin1981quantized, thouless1982quantized}.
A simplified two degrees of freedom (DOF) model, proposed as a description of spinless polyacetyline by Su, Schrieffer and Heeger \cite{su1979solitons},  known as the SSH model, displays a topologically non-trivial band gap characterized by a non-zero Zak phase, and as such it admits protected edge states \cite{shen2012topological}. Higher-order topological insulators (HOTI) extend this bulk--boundary correspondence: an ($N$)-dimensional system may host topological states on boundaries of dimension (${N-1}$) or lower, such as zero-dimensional corner states and one-dimensional edge states in two-dimensional systems \cite{benalcazar2017quantized,ezawa2018higher}. 
Although these effects were first found in quantum systems, they can be generalized to systems of linear, undamped oscillators, since their motion can be described by the Shr{\"o}dinger equation \cite{susstrunk2016classification, huber2016topological}. This equivalence allows to reproduce analogues of quantum physical phenomena in macroscopic mechanically designed structures, in which geometry and elastic behaviour can be controlled to create, tune, and experimentally visualize localized modes.  Topological effects have been observed in mechanical \cite{susstrunk2016classification, vila2017observation, wakao2020higher, duan2023numerical, zhang2023higher, duan2024weak, gao2025simplified, huber2016topological, chaplain2020topological, Haslinger2022, chaplain2023tunable, huang2023topological, ma2024observation, chen2018study, jara2022topological, miranda2024mechanical, ozdemir2024rigorous, pal2017edge, dal2024topological, morvaridi2024tunable,Garau2019,ongaro2025closed, serra2018observation,Carta2026,fan2019elastic,wakao2020higher}, acoustic \cite{chen2018acoustic, zheng2019observation, fleury2016floquet, liu2019trapping, xue2019acoustic, xiong2025experimental, PhysRevLett.128.116802, Xue2022, Zhu2023} and optical \cite{lu2014topological, liu2019photonic, yves2017crystalline, smirnova2020nonlinear} systems, as well as in electronic circuits \cite{liu2019topologically}.


Previous studies of mechanical topological insulators have mainly followed two complementary strategies. Simplified mass–spring models provide a transparent route to band topology and closed-form topological invariants \cite{ongaro2025closed, wakao2020higher, zhang2023higher}, but they generally do not account for beam bending, rotational degrees of freedom, ligament eccentricity, or the manufacturability constraints of continuous elastic structures. Conversely, full finite-element simulations and experiments capture realistic geometries and deformation fields \cite{duan2024weak, duan2023numerical, vila2017observation, serra2018observation,fan2019elastic}, but they are computationally demanding and less suited to systematic phase-space exploration. 
Serra-Garcia et al. \cite{serra2018observation} experimentally demonstrated a phononic quadrupole insulator based on coupled out-of-plane plate resonances. Fan et al. \cite{fan2019elastic} investigated higher-order topological states in a continuous elastic honeycomb lattice through numerical simulations and experiments on out-of-plane vibrations. Wakao et al. \cite{wakao2020higher} considered vectorial in-plane dynamics of a mass-spring breathing kagome lattice, where coupling between longitudinal and transverse motion affects its higher-order topological phase. Together, these works provide a range of simplified and continuous systems, motivating
the development of a reduced continuum-informed model that can bridge the gap between lattice descriptions, which preserve efficiency and interpretability, and full simulations, which retain the structural parameters that control the band gap, the topological transition, and the resulting boundary-localized modes. Here, we develop such a model — mechanically faithful and analytically tractable — capable of predicting topology, dispersion, boundary modes, and experimental response within a single framework. 

To do this, we derive a 12-degree-of-freedom Hermitian dynamical matrix for a continuous beam-connected lattice of rigid squares. The model retains translational and rotational degrees of freedom and includes ligament eccentricity. We use it to design a 2D SSH phononic crystal, characterize its bulk polarization, identify a geometry-related topological transition, and predict the emergence of edge and corner modes. We validate the analytical results using finite element simulations and experimental measurements on 3D-printed samples. 

The centered configuration serves as the principal reference case, while selected off-centered configurations are investigated to demonstrate how the model captures eccentricity-dependent changes in bulk topology and boundary localization. However, the main aim is to establish and validate the predictive possibilities of the modeling approach.

\section{Model}\label{sec:Model}

We consider a phononic crystal whose unit cell (UC) consists of four rigid squares of side length $2a$, mass $m$ and moment of inertia $I$, each having three degrees of freedom (DOF), two translational (in the $x$ and $y$ directions) and one rotational, totalling 12 DOFs. We indicate the three DOFs of the four square subunits $\alpha=A,B,C,D$ within the $(p,q)$ unit cell as 
\begin{align}
    &u_{p,q}^\alpha\text{: $x$ displacement,}\nonumber\\
    &v_{p,q}^\alpha\text{: $y$ displacement,}\\
    &\theta_{p,q}^\alpha\text{: counterclockwise rotation.}\nonumber
\end{align}

\begin{figure}[ht]
\centering
\begin{subfigure}{\linewidth}
    \centering
    \captionsetup{justification=raggedright,singlelinecheck=false, margin=10mm}
    \caption{}
    \includegraphics[width=0.8\linewidth]{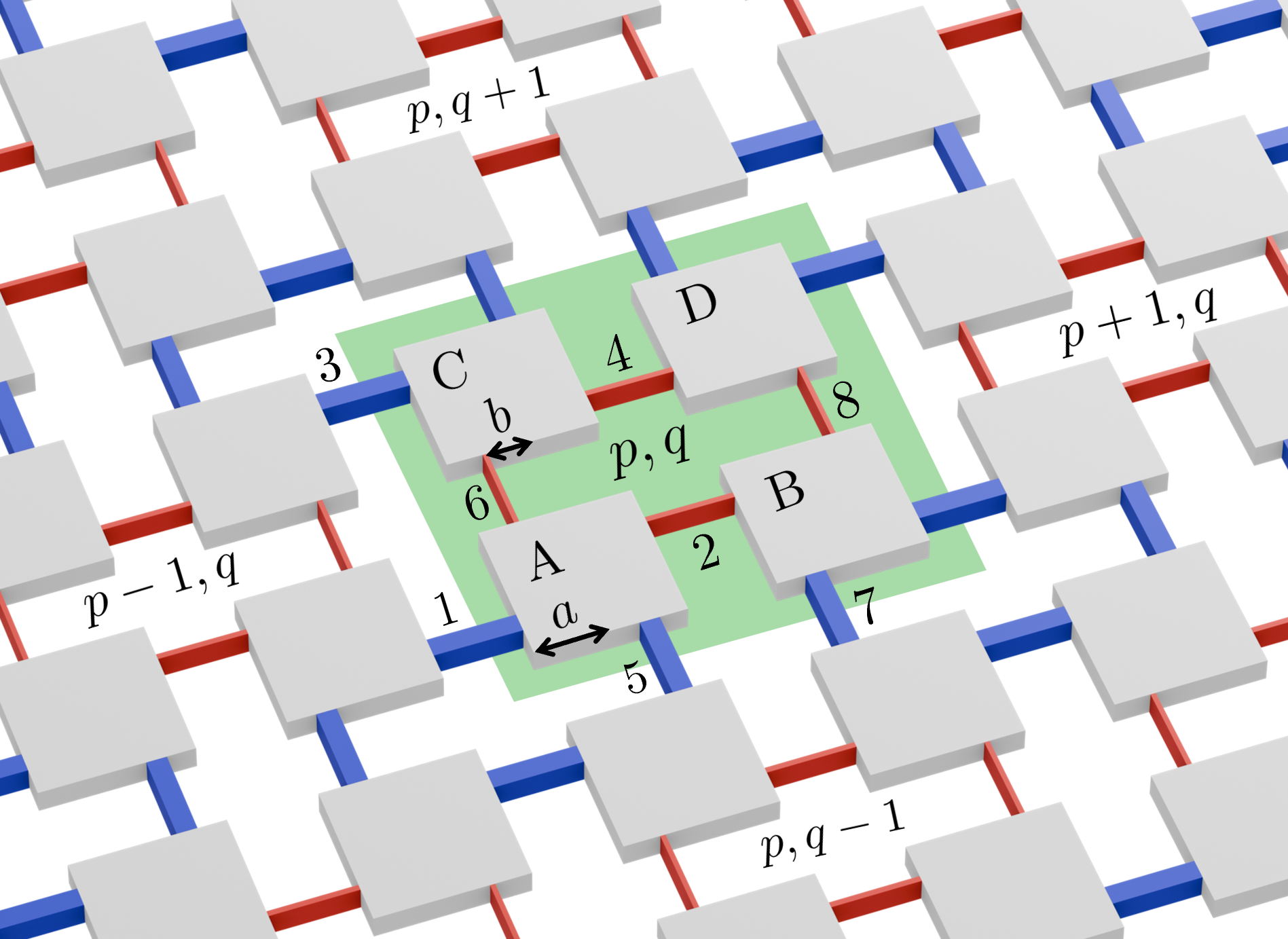}
    \label{fig:uc}
\end{subfigure}
\vspace{-2mm}

\begin{subfigure}{\linewidth}
   \centering
    \captionsetup{justification=raggedright,singlelinecheck=false, margin=10mm}
    \caption{}
    \includegraphics[width=0.7\linewidth]{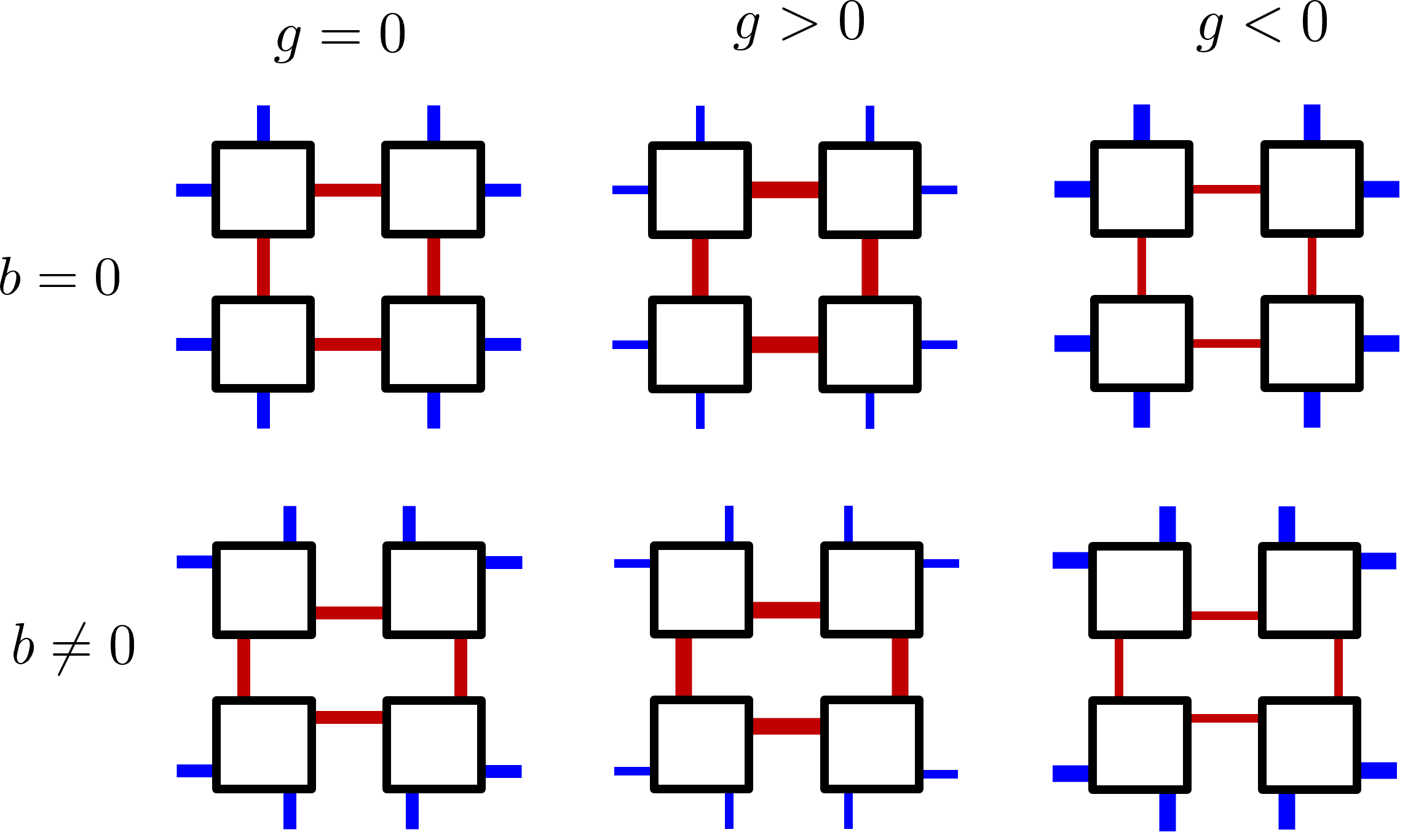}
    \label{fig:uc_g_b}
\end{subfigure}

\caption{(a) Geometry of the phononic crystal showing unit cell (UC) in green, with labeled squares, neighboring cells, and beam numbering. (b) Effect of the parameters $g$ and $b$ on the geometry of the UC.}
\label{fig:UC_geometry}
\end{figure}

The square subunits are connected by eight rectangular ligaments of width $s_i$ and length $l$, that can be positioned off-center using an eccentricity factor $b$ (see Fig. \ref{fig:UC_geometry}). The structure has constant thickness $h$. With this type of simple unit cell, whose geometrical characteristics are defined parametrically, it is possible to build a reduced order model that reproduces the dynamical behaviour of  various types of real phononic crystals with complex geometries.  

Assuming that the ligaments are thin, namely that their cross-section is significantly smaller than their length, they can be modelled as Euler-Bernoulli beams. Under the hypothesis of small displacements and rotations, the forces and moments acting upon a given mass are derived in Appendix \ref{app:Forces-and-Moments}. Using these expressions of forces and moments, we
obtain a set of $12$ coupled linear equations of motion. The system is Hermitian and can be made dimensionless by substituting ${\displaystyle{\overline{u}=u/a}}$, ${\displaystyle{\overline{v}=v/a}}$ and $\displaystyle{\tilde{\theta}=\theta\sqrt{2/3}}$. Defining the state vector of the UC $(p,q)$ as 
$\displaystyle{\ket{\psi_{p,q}}=\nonumber(\overline{u}_{p,q}^A, \overline{u}_{p,q}^B, \overline{u}_{p,q}^C, \overline{u}_{p,q}^D, \overline{v}_{p,q}^A, \overline{v}_{p,q}^B, \overline{v}_{p,q}^C, \overline{v}_{p,q}^D, \tilde{\theta}_{p,q}^A, \tilde{\theta}_{p,q}^B, \tilde{\theta}_{p,q}^C, \tilde{\theta}_{p,q}^D)^T}$, 
we assume a solution in the form of Bloch waves such that
\begin{align}\label{eq:Bloch_form_analytical}
    \displaystyle{\ket{\psi_{p+n_x,q+n_y}}=e^{i\left[\left(k_xn_x+k_yn_y\right)d-\omega t\right]}\ket{\psi_{p,q}}},
\end{align}
where $k_x$ and $k_y$ are the $x$ and $y$ components of the wave vector, $\omega$ is the angular frequency of the Bloch wave and $d=2\left(2a+l\right)$ is the lattice constant. Thus, the system of equations of motion in the time-harmonic regime can be written as the following eigenvalue problem:
\begin{align}\label{eq:EVP}
    H\ket{\psi}=\Omega^2\ket{\psi},
\end{align}
where $\Omega=\omega \sqrt{ml/(EA)}$ are the dimensionless eigenfrequencies and $H$ is the $\displaystyle{12\times 12}$ stiffness matrix.

 
We exploit the high generality of the model to build a mechanical 2D Su–Schrieffer–Heeger (SSH) system, alternating the width of the beam in the $x$ and $y$ directions. In this way, we create a distinction between intracell beams connecting squares within a UC, and intercell beams connecting UCs to each other. The system is shown in Fig. \ref{fig:uc}, where intracell and intercell beams are represented in red and blue, respectively. Using the beam numbering of Fig. \ref{fig:uc}, the width of intracell and intercell beams is $s_{\text{int}}$ for beams $2,4,6,8$ and $s_{\text{ext}}$ for beams $1,3,5,7$. Their expression is
\begin{align}\label{eq:beam_width}
    s_{\text{int,ext}}=\overline{s}(1\pm g/2),
\end{align}
where the parameter $g$ denotes the relative width difference with respect to the average width $\overline{s}$, as represented in Fig \ref{fig:uc_g_b}. With this description, $g=0$ corresponds to the ``monoatomic" case for which all beams have the same width. For $g<0$, the intracell beams are thinner than the intercell beams, and viceversa for $g>0$.

\begin{figure}[!ht]
\centering

\begin{subfigure}{\linewidth}
    \centering
    \caption{}
    \hspace*{0mm}
    \includegraphics[width=1\linewidth]{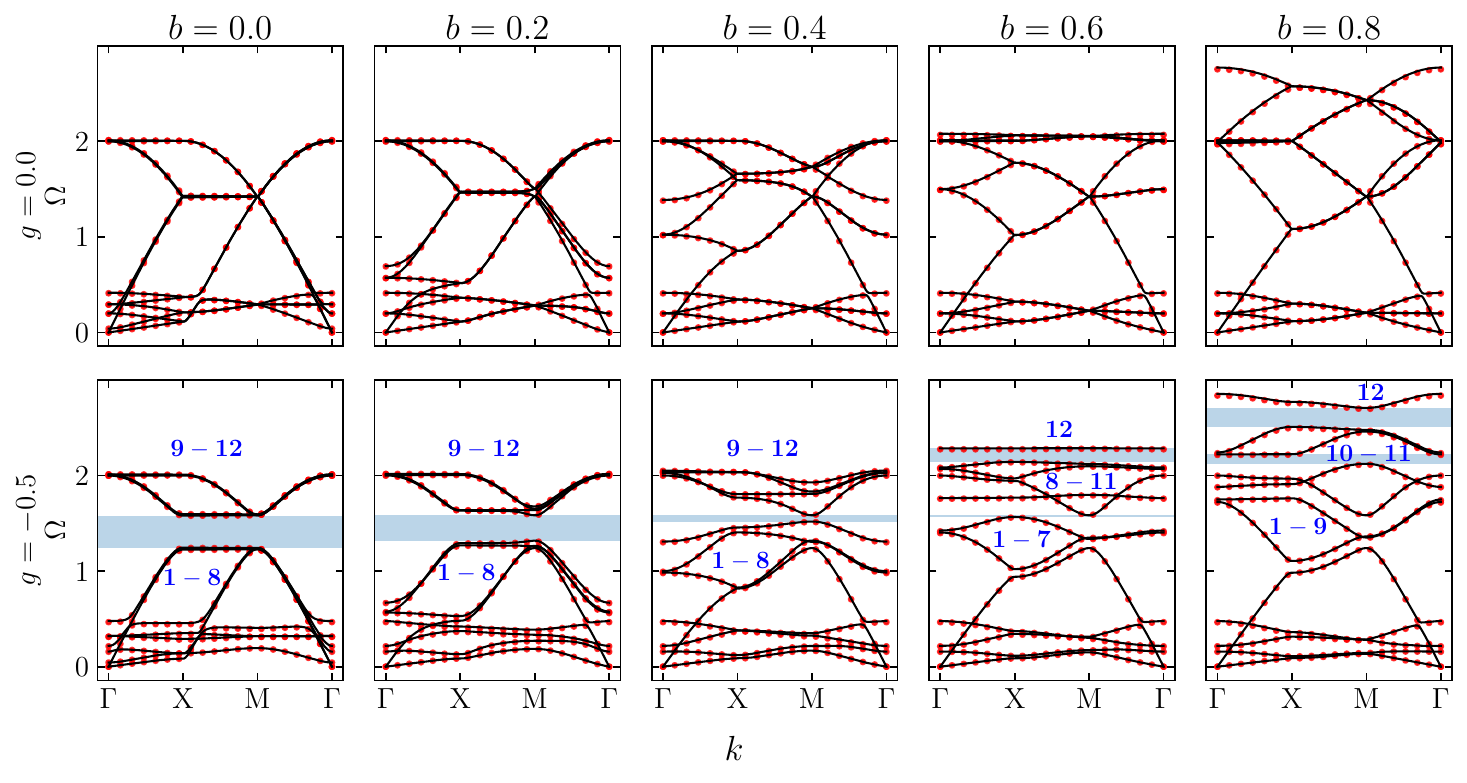}
    \label{fig:dr}
\end{subfigure}

\vspace{-10mm} 

\begin{subfigure}{\linewidth}
    \centering
    \caption{}
    \hspace*{4mm}
    \includegraphics[width=\linewidth]{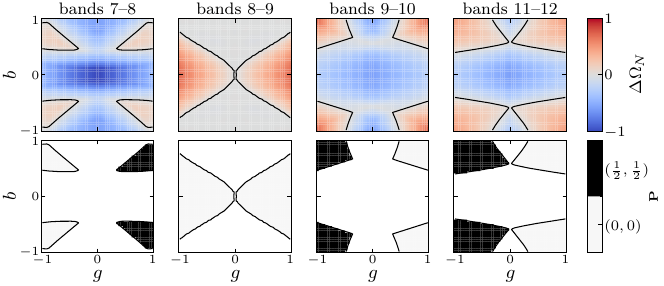}
    \label{fig:gaps_and_polarization}
\end{subfigure}

\caption{
(a) Dispersion diagrams obtained from the analytical calculations (solid black lines) and the FEM model (red dots), for the monoatomic ($g=0$) and dimerized ($g\neq0$) cases, and for different off-center positions $b$. The band gaps are shown by the light blue areas, and band indices of groups of branches are reported in blue. (b) Top: Map of complete band gap width for varying $b$ and $g$, between bands $7$--$8$, $8$--$9$, $9$--$10$ and $11$--$12$, where $\Delta\Omega_N$ represents the difference between the minimum frequency of band $N+1$ and the maximum frequency of band $N$. The red area with $\Delta\Omega_N >0$ indicates the presence of a band gap between bands $N$ and $N+1$. Bottom: Vector bulk polarization of each band gap.
}
\label{fig:periodic}
\end{figure}

\begin{figure*}[t]
    \centering
    \captionsetup[subfigure]{justification=centering, singlelinecheck=false}

    \begin{subfigure}[c]{0.32\textwidth}
    \vspace{15mm}
        \centering
        \caption{}
        \includegraphics[width=1.25\linewidth]{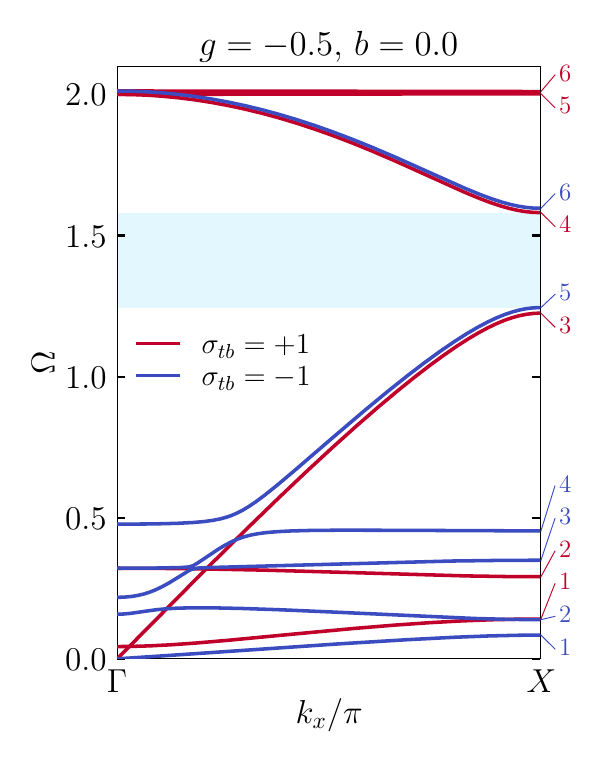}
        \label{fig:dr_odd_even}
    \end{subfigure}
    \hfill
    \begin{minipage}[c]{0.65\textwidth}
        \centering

        \begin{subfigure}[c]{\linewidth}
            \centering
            \caption{}
            \includegraphics[width=0.7\linewidth]{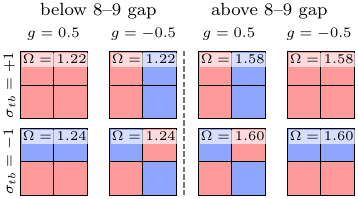}
            \label{fig:mode_inversion}
        \end{subfigure}

        \vspace{0mm}

        \begin{subfigure}[c]{\linewidth}
            \centering
            \caption{}
            \includegraphics[width=0.7\linewidth]{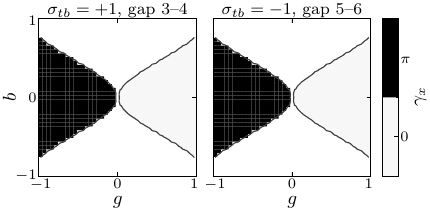}
            \label{fig:sector_zak_map}
        \end{subfigure}

    \end{minipage}

    \caption{Mirror-resolved interpretation of the 8--9 band gap.
    (a) Dispersion relation along \(\Gamma-X\), with bands colored by the
   mirror eigenvalue \(\sigma_{tb}=\pm 1\).
    (b) Longitudinal mode character at \(X\), showing the inversion of the
    acoustic- and optical-like modes in both mirror sectors.
    (c) Sector-resolved Zak phase maps for the two relevant mirror sectors,
    masked by the complete 8--9 bulk band gap.
    }
    \label{fig:mirror_sector_89}
\end{figure*}

The model also allows the off-center position of the beams to be varied, as quantified by the parameter \(b\). Its effect on the unit cell is shown in Fig.~\ref{fig:uc_g_b}. For \(b=0\), the unit cell is centered and has \(C_{4v}\) symmetry, including fourfold rotational symmetry and mirror-reflection symmetries. For \(b\neq0\), the off-center beam attachments break the fourfold rotational symmetry while preserving the horizontal and vertical mirror symmetries. The unit cell therefore has \(C_{2v}\) symmetry, with only a twofold rotation and two mirror reflections.

\begin{table}[h]
\begin{tabular}{|l|l|}
\hline
Material                  & Aluminium   \\ \hline
Density $\rho$ (kg$/$m$^3$) & $2700$ \\ \hline
Young's modulus $E$ (GPa)                         & $70$                         \\ \hline
Thickness $h$ (mm)                                     & $1$                         \\ \hline
Square length $2a$ (mm)                                & $50$                     \\  \hline
Beam length $l$ (mm)                                  & $10$                        \\ \hline
Average beam width $\overline{s}$ (mm)                             & $1$                 \\ \hline
\end{tabular}
\caption{Material and geometrical parameters of the analytical and numerical models. The widths of the thinner and thicker beams are evaluated as in (\ref{eq:beam_width}).}
\label{tab:an}
\end{table}

The dispersion diagrams for different values of $g$ and $b$ are shown in Fig. \ref{fig:dr}, along with the corresponding band gaps. The material and geometrical parameters used in the computations are specified in Table \ref{tab:an}. In the  monoatomic case $(g=0)$, the spectrum remains gapless for all $b$. For $b=0$, the lattice has $C_{4v}$ symmetry. The fourfold rotation symmetry makes the two
principal directions equivalent, so several modes become degenerate along the high-symmetry path,
forming a highly degenerate nodal manifold. When $b\neq 0$, the symmetry is lowered from $C_{4v}$ to $C_{2v}$. As a result, modes that were equivalent under \(90^\circ\) rotations are no longer symmetry-related, and the high-order degeneracy is lifted. The remaining \(C_{2v}\) symmetry still relates pairs of modes, so the spectrum separates into nearly degenerate pairs. In addition, a large eccentricity shifts several branches to higher frequencies. This shift follows from the off-center ligament attachment, which enhances the translation–rotation coupling and introduces an additional term in the effective rotational stiffness. In particular, Eq.~\eqref{eq:Forces_moments_F2_to_F5} contains the contribution \(M_5=F_5b=(EA_2/l)b^2\theta^{\mathrm B}_{p,q}\), showing explicitly that the rotational stiffness grows with \(b^2\).

When dimerization $(g\neq 0)$ is introduced, a full band gap opens around the nodal frequency, between bands $8$ and $9$. The gap originates from the SSH-type splitting of the longitudinal acoustic- and optical-like modes. At the same time, two fourfold-degenerate nodal manifolds remain below and above the gap. Increasing $b$ again lifts these degeneracies and shifts the bands in frequency. As a result, the 8--9 gap closes near $b\simeq0.4$ in the plotted parameter range. Figure~\ref{fig:gaps_and_polarization} further shows that, for large eccentricity and strong dimerization, additional complete gaps open in the upper part of the spectrum, notably between bands $7$--$8$, $9$--$10$, and $11$--$12$. 

Unlike scalar or mass--spring mechanical SSH systems, the present model incorporates ligament bending and translation--rotation coupling. These effects introduce richer physics, including additional stiffness channels and multiband hybridization, while mantaining a more realistic description of fabricated samples.

Many topological insulators (TI) are characterized by the Chern number, which is the integral of the Berry curvature $\mathcal{F}$ over the Brillouin zone (BZ)
to determine a non-trivial topological transition \cite{kane2005quantum}. 
However, our system is characterized by inversion symmetry which implies
${\mathcal{F}(-\mathbf{k})=\mathcal{F}(\mathbf{k})}$, whereas time-reversal symmetry requires \({\mathcal{F}(-\mathbf{k})=-\mathcal{F}(\mathbf{k})}\). Therefore, the Berry curvature vanishes for every  $\mathbf{k}$ in the BZ, where the band is isolated and non-degenerate \cite{fu2011topological, liu2017novel, liu2018topological}. For such TIs, the presence of a topological phase transition is encoded in the 2D bulk polarization $\mathbf{P}=(P_x,P_y)$.  
The bulk polarization is related to the vectored Zak phase
\(\mathbf Z=(Z_x,Z_y)\) as
\begin{align}
    P_j=\frac{Z_j}{2\pi}\quad \mathrm{mod}\;1,
    \qquad
    Z_j =
    i\sum_{n=1}^{N_{\rm occ}}
    \int_{0}^{2\pi/d}
    \left\langle
    \psi_n(\mathbf{k})
    \middle|
    \partial_{k_j}
    \psi_n(\mathbf{k})
    \right\rangle
    dk_j ,
\end{align}
 where $j=x,y$ and $N_{occ}$ is the set of bands below the band gap of interest. Given the inversion symmetry of the system, the bulk polarization vector is connected to the parity of the eigenstates at the high symmetry points, meaning that  its components can be calculated as follows \cite{
 PhysRevB.100.075437}: 
\begin{align}
    P_j=\frac{1}{2}\left[\frac{-i}{\pi}\text{ln}\left(\frac{\det\left[\mathcal{B}\left(K_j\right)\right]}{\det\left[\mathcal{B}\left(\Gamma\right)\right]}\right)\;\text{mod }2\right],
\end{align}
where 
\begin{align}
    \left[\mathcal{B}\left(K_j\right)\right]_{m,n}=\left\langle \psi_m(K_j) \right\vert \hat{C}_2 \left\vert \psi_n(K_j) \right\rangle
\end{align}
is the so-called sewing matrix \cite{fang2012bulk} and $m,n=1,\ldots,N_{\rm occ}$. In addition, ${K_x=X}$ and ${K_y=Y}$ correspond to the high symmetry points ${\mathbf{k}=(\pi/d,0)}$ and ${\mathbf{k}=(0,\pi/d)}$ respectively, and $\hat{C}_2$ is the twofold rotation operator of the system which acts on the DOFs as:
\begin{align}
    \hat{C}_2: \begin{cases}
        u_A\leftrightarrow -u_D,\quad v_A\leftrightarrow -v_D,\quad \theta_A\leftrightarrow\theta_D,\\
        u_B\leftrightarrow -u_C,\quad v_B,\leftrightarrow -v_C,\quad \theta_B\leftrightarrow\theta_C. \\
    \end{cases}
\end{align}

The bulk polarization of the phononic crystal has been computed for all $g$ and $b$ values and mapped in Fig. \ref{fig:gaps_and_polarization}. As discussed below, in the symmetrical case {($b=0$)} results reveal topological phase transitions from trivial to non-trivial at $g=0$ (bands 8-9). In other cases, no continuous transition is observed. Gaps between bands $9$--$10$ and $11$--$12$ are non-trivial for $g<0$, whereas the gap between bands $7$--$8$ is non-trivial for $g>0$. This inversion reflects the multiband nature of the beam lattice, where translation--rotation coupling hybridizes the bands and can
reverse the parity inversion associated with a given gap. The finite-spectrum of the higher frequency bands and the corresponding modes are \textcolor{black}{discussed in Sec. \ref{sec:localized_modes}}.

The gap between bands $8$--$9$ has a trivial cumulative bulk polarization, \(\mathbf P=(0,0)\), despite the presence of a SSH-type band inversion. This apparent discrepancy arises from the multiband structure of the spectrum. 
\textcolor{black}{Near the $X$ point, bands $7$ and $8$ lie below the band gap and bands $9$ and $10$ lie above. The corresponding modes are longitudinal since the displacement of the four squares is in the $x$ direction. The mode character at the $X$ point for these bands is shown in \ref{fig:mode_inversion}. Each pair contains a symmetric and an antisymmetric mode in the $y$ direction, at an almost identical frequency. For $g>0$, bands $7$ and $8$ correspond to acoustic-like modes, since they are symmetric in $x$, while bands $9$ and $10$ correspond to optical-like modes since they are antisymmetric in $x$. On the contrary, for $g<0$, the lower bands $7$ and $8$ correspond to optical-like modes while the higher bands $9$ and $10$ correspond to acoustic-like modes in $x$. This inversion of parity is a characteristic of SSH-like systems below the non-trivial band gap. In this case, it takes place for a pair of bands instead of a single band, due to the system being quasi-2D on the $\Gamma-X$ path, and having top-bottom mirror symmetry. In fact, due to the mirror symmetry, the band structure can be divided in two independent sectors, where the modes are classified by $\sigma_{tb}=+1$ or $\sigma_{tb}=-1$ depending if they are symmetric or antisymmetric in the $y$ direction, where $\sigma_{tb}$ are the eigenvalues of the top-bottom mirror reflection operator. The separation into symmetry sectors is detailed in Appendix \ref{sec:separation_odd_even}, and Fig \ref{fig:dr_odd_even} shows the dispersion diagram with the bands belonging to ${\sigma_{tb}=+1}$ and ${\sigma_{tb}=-1}$ colored in red and blue respectively. In fact, we clearly see two symmetric-antisymmetric pairs of bands which are almost coincident.} 
The two non-trivial contributions therefore cancel in the cumulative polarization of the full subspace below the 8--9 gap, giving a trivial total invariant. However, the topology remains visible at the symmetry-sector level: computing the Zak phase of each sector as 
\begin{align}
    \gamma_{x}
    =
    i\sum_{n=1}^{N_{\text{occ}}}
    \int_{0}^{2\pi/d}
    \left\langle
    \psi_{n,\pm}(k_x)
    \middle|
    \partial_{k_x}
    \psi_{n,\pm}(k_x)
    \right\rangle
    dk_x ,
\end{align}
where \(N_{\text{occ}}\) is the number of bands below the relevant sector gap, and the $\pm$ index indicates the state with positive or negative symmetry eigenvalue. Fig. \ref{fig:sector_zak_map} shows how for $g<0$ the system has a non-trivial Zak phase, with a topological phase transition occurring at $g=0$ and $b=0$.

The same argument applies along the \(\Gamma-Y\) direction, where the left-right mirror symmetry separates the spectrum into independent \(\sigma_{lr}=\pm1\) sectors. Thus, the 8--9 gap is characterized not by a nontrivial cumulative polarization of all bands below the gap, but by sector-resolved SSH topology along both principal directions. This sector topology gives rise to four ribbon edge states, shown in Fig.~\ref{fig:ribbon_displacement}, and to eight corner-localized modes in finite samples, shown in Fig.~\ref{fig:finite_maps}.

Since the sign of \(g\) determines which set of ligaments, intracell or intercell, is stiffer, changing the sign of \(g\) therefore exchanges the strong and weak beam connections, in analogy with the hopping inversion of the SSH chain. In this context, for \(g>0\), the dominant stiffness bonds lie inside the chosen unit cell, so the localized Wannier-like vibration centers are contained within the cell and the corresponding Zak phase is trivial. For $g<0$, the dominant stiffness bonds connect neighbouring unit cells. The Wannier-like centers are then displaced to the unit-cell boundary, which gives a Zak phase of \(\pi\) along that direction \cite{king1993theory, resta1994macroscopic, PhysRevB.100.075437}.

In the 8--9 gap, this mechanical SSH-type inversion occurs independently in the two mirror sectors of the longitudinal modes, which acquire a nontrivial Zak phase. When the lattice is cut through the Wannier-like center, it leaves edge-localized ribbon bands inside the bulk band gap. In a finite system, these edge bands constitute effective one-dimensional SSH boundary channels, so an edge supports localized end states. Since the ends of the one-dimensional edges coincide with the corners of the finite lattice, these end states appear as corner-localized modes.

We validate the model with finite element method (FEM) simulations of an equivalent system in 2D, plane stress. The simulations perform an eigenfrequency analysis of a linear elastic and isotropic continuum, with Floquet-Bloch conditions applied on the boundaries of the UC.

The mesh is chosen so that there are at least two elements along the width of the smallest beams. A finer mesh has a negligible effect on the calculated results for the dispersion relation. Figure \ref{fig:dr} shows how for the tested cases, there is an excellent 
match between the analytical solutions and the FEM simulations, highlighting the accuracy of the mass-beam system discretization. However, analytical calculations reduce the computation time by various orders of magnitude compared to the FEM simulations.

\section{Localized topological modes}\label{sec:localized_modes}
\subsection{Ribbon}

\begin{figure}[!h]
  \centering

  \begin{subfigure}[t]{\linewidth}
    \centering
    \caption{}
    \label{fig:Ribbon_UC}
    \resizebox{\linewidth}{!}{%
      \includegraphics[]{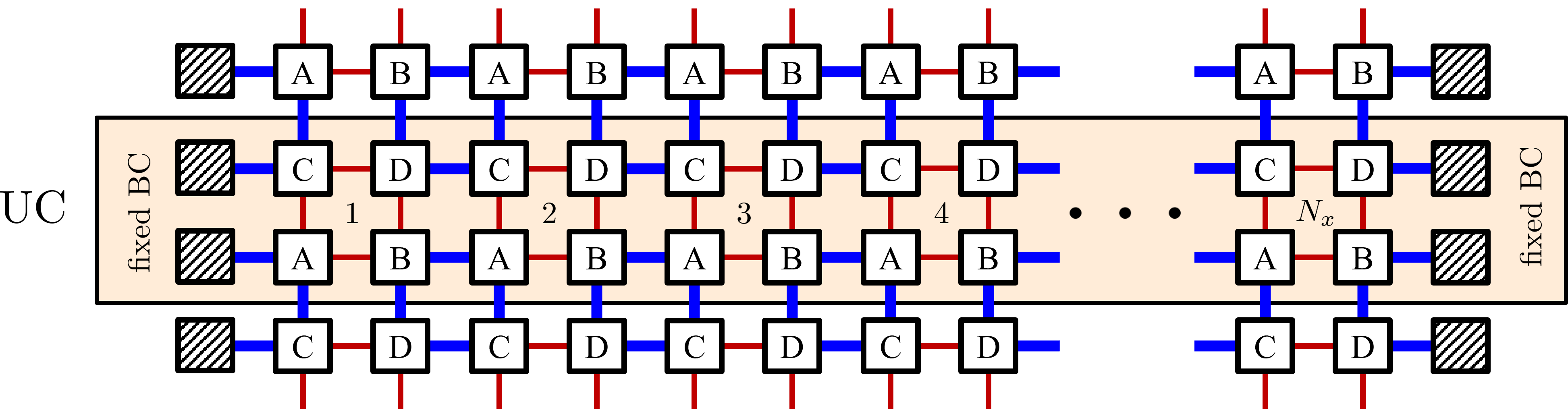}%
    }
  \end{subfigure}

  \vspace{0mm}

  \begin{subfigure}[t]{\linewidth}
    \centering
    \caption{}
    \label{fig:Ribbon_DR}
    \resizebox{\linewidth}{!}{%
      \includegraphics[]{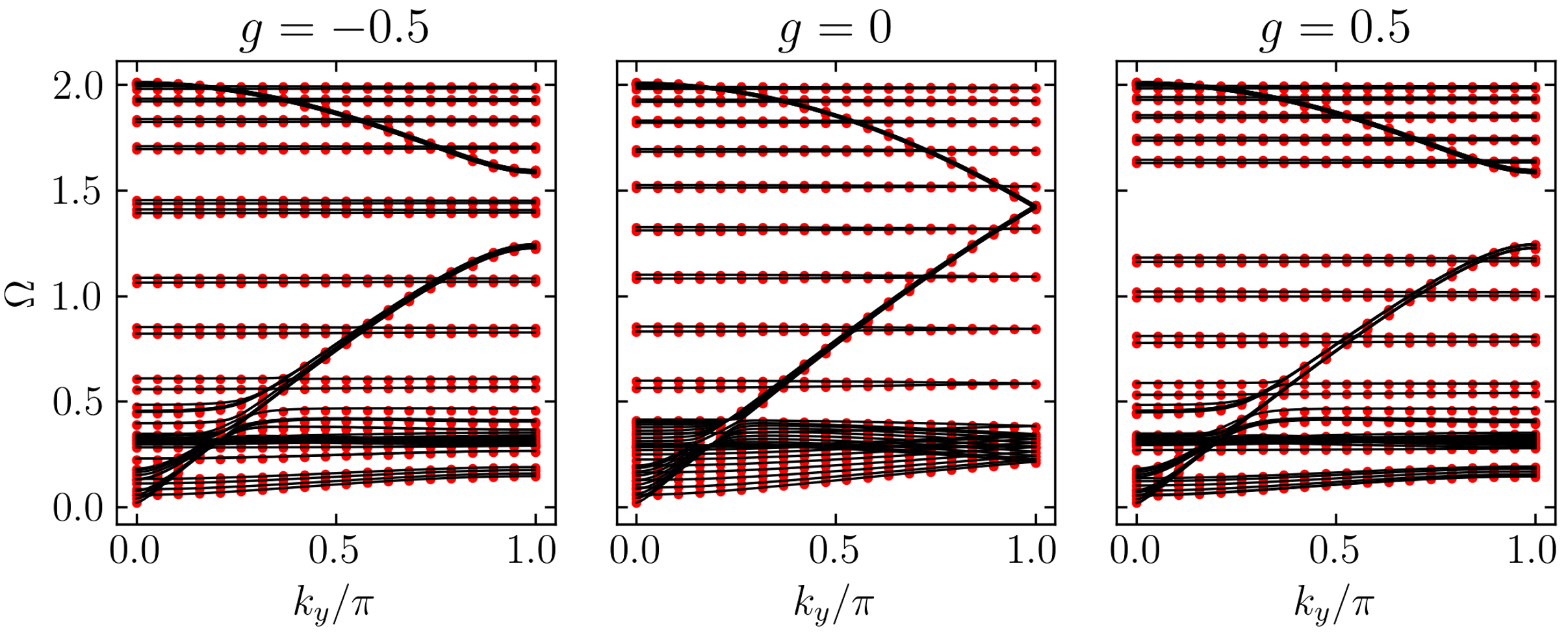}%
    }
  \end{subfigure}

\vspace{-2mm}

\begin{subfigure}[t]{0.49\linewidth}
  \centering
  \caption{}
  \label{fig:Omega_varying_g}
  \includegraphics[width=\linewidth]{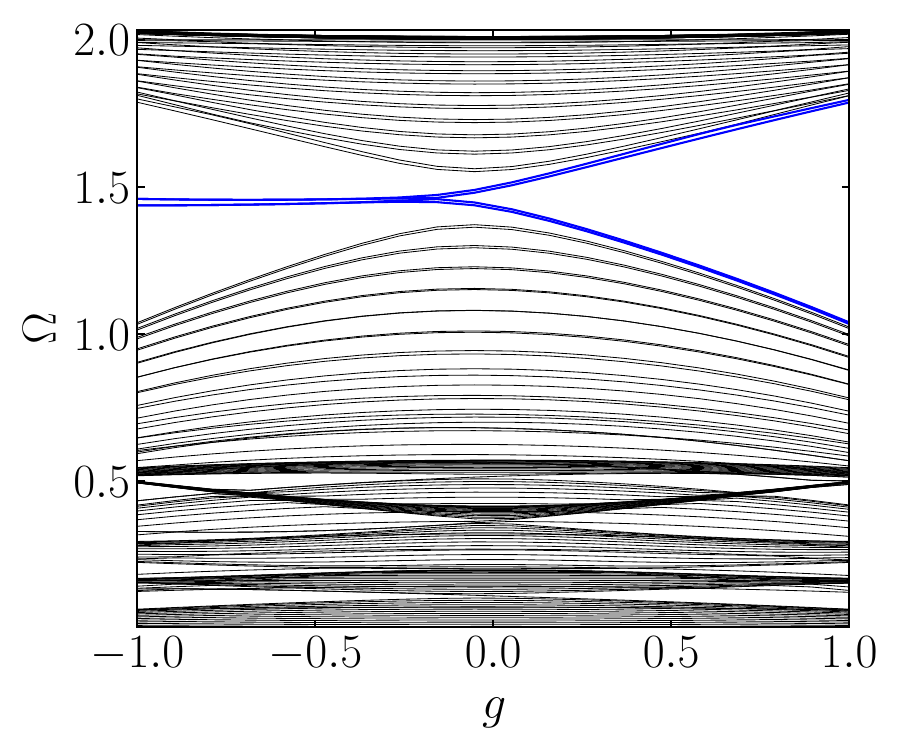}
\end{subfigure}
\hfill
\begin{subfigure}[t]{0.49\linewidth} 
\caption{} 
\label{fig:ribbon_displacement} 
\includegraphics[width=1\linewidth]{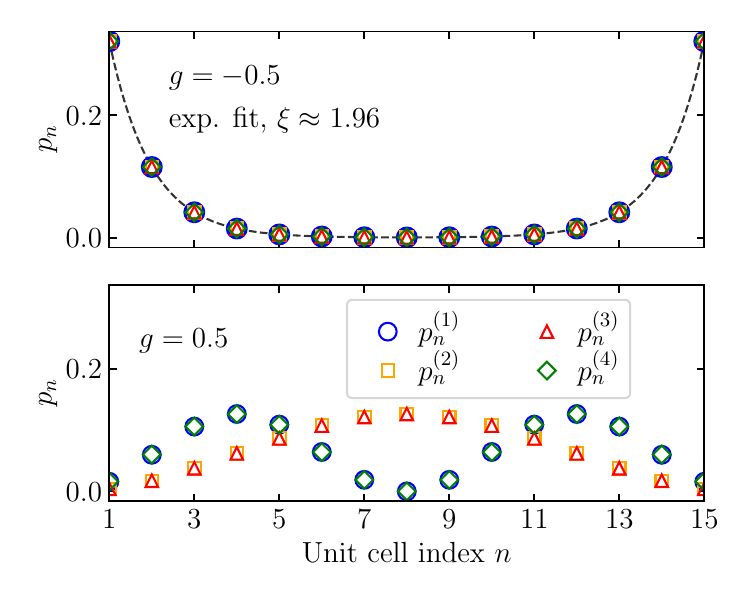}
\end{subfigure}
\vspace{-5mm}
  \caption{(a)
         Ribbon macrocell, Floquet-Bloch conditions have been applied between the top and the bottom of the ribbon unit. (b) Spectrum of the ribbon with $N_x=5$ versus $k_y$ for the non-trivial, gapless and trivial cases, obtained from the analytical calculations (solid black lines) and the FEM model (red dots). (c) Eigenfrequencies for varying $g$. $N_x=15$ UCs were used for a better visualization of the exponential decay. (d) Spatial profile $p_n$ of the four ribbon edge modes. Dashed black curve indicates two-sided exponential fit used to estimate the localization length $\xi$.}
  \label{fig:Displacement_ribbon}
\end{figure}

TIs obey bulk-edge correspondence, whereby a non-trivial system exhibits edge states inside the BG
\cite{chen2020elementary, xiong2025experimental}. In higher order topological insulators (HOTIs), these can appear in the form of edge or corner modes, which are of particular interest for energy localization and harvesting. To observe edge states, we study a system in the form of a ribbon, periodic in the $y$ direction and containing $N_x$ UCs in the $x$ direction, with fixed boundary conditions at the left and right edges (Fig. \ref{fig:Ribbon_UC}). For illustration purposes, here we consider \(b=0\). A similar off-centre system with \(b=0.8\) is illustrated in Appendix \ref{sec:app:ribbon}.

\FloatBarrier

The ribbon eigenfrequencies obtained from the analytical calculations are in excellent agreement with the FEM simulations across the entire Brillouin zone for all systems, including the edge modes (see Fig. \ref{fig:Ribbon_DR}).
The ribbon built using topologically non-trivial UCs ($g<0$) exhibits four modes inside the band gap, see Fig. \ref{fig:Ribbon_DR} (left). The corresponding eigenmodes are localized on the edges of the system, see Fig. \ref{fig:ribbon_displacement} (top), in agreement with bulk-edge correspondence of topologically non-trivial systems. The four nearly degenerate edge states arise from the mirror symmetry of the system; for each boundary, there is an edge state for the even and one for the odd sector band gap. Conversely, the trivial system ($g>0$) shows a fully gapped spectrum, see Fig. \ref{fig:Ribbon_DR} (right), delimited by bulk states, see Fig. \ref{fig:ribbon_displacement} (bottom). When varying continuously the parameter $g$ (Fig. \ref{fig:Omega_varying_g}), the two edge modes shown in blue shift from inside the band gap (at $\Omega\approx 1.4$) for $g<0$, to join progressively the bulk states at $g>0$, demonstrating a clear phase transition at $g=0$.

 The spatial profile of each ribbon mode is quantified by the unit-cell-resolved modal weight
\begin{equation}
    p_n^{(m)}=\sum_{\alpha=1}^{12}\left|\psi_{n,\alpha}^{(m)}\right|^2,
\end{equation}
where \(n\) labels the unit cell across the ribbon, \(\alpha\) runs over the 12 DOFs of each unit cell, and \(m=1,\ldots,4\) labels the localized modes. The weights are normalized such that \(\sum_n p_n^{(m)}=1\). Since the finite ribbon supports modes localized at both boundaries, the numerical profiles are fitted with a two-sided exponential envelope,
\begin{equation}
    p_{\mathrm{fit}}^{(m)}(x_n) =
    A_L e^{-2x_n/\xi}
    + A_R e^{-2(L-x_n)/\xi}
    + c,
\end{equation}
where \(\xi\) is the localization length, \(L\) is the ribbon width, and \(A_L\) and \(A_R\) account for the modal weight localized at the left and right boundaries, respectively. The constant offset \(c\) accounts for a small residual bulk contribution or a numerical floor in the fit. Performing the fit gives \(\xi=1.96\), corresponding to a decay length of approximately two unit cells from each edge. The modal profiles and their exponential fits are shown in Fig.~\ref{fig:ribbon_displacement}.
For an ideal SSH chain, the edge-state amplitude decays as
\begin{equation}
    |\psi_n| \propto \left|\frac{k_{\rm weak}}{k_{\rm strong}}\right|^n,
\end{equation}
giving the localization length
\begin{equation}\label{eq:SSH_localization}
    \xi_{\rm SSH}
    =
    \frac{1}{\ln\left(k_{\rm strong}/k_{\rm weak}\right)}
\end{equation}
in units of the lattice period~\cite{su1979solitons,asboth2015short}. In our case, using \(g=-0.5\) in Eq.~\eqref{eq:beam_width}, the ratio between the strong and weak beam widths is \(5/3\). Therefore, Eq.~\eqref{eq:SSH_localization} gives \(\xi_{\rm SSH}=1.96\), in agreement with the localization length extracted from the exponential fit. Although the present mechanical model contains additional internal degrees of freedom, this SSH estimate provides a useful reference scale for the numerically fitted decay length. 

\subsection{Finite system}

\begin{figure}[t]
    \centering


    \begin{subfigure}[t]{0.32\linewidth}
        \centering
        \caption{}
        \includegraphics[width=\linewidth]{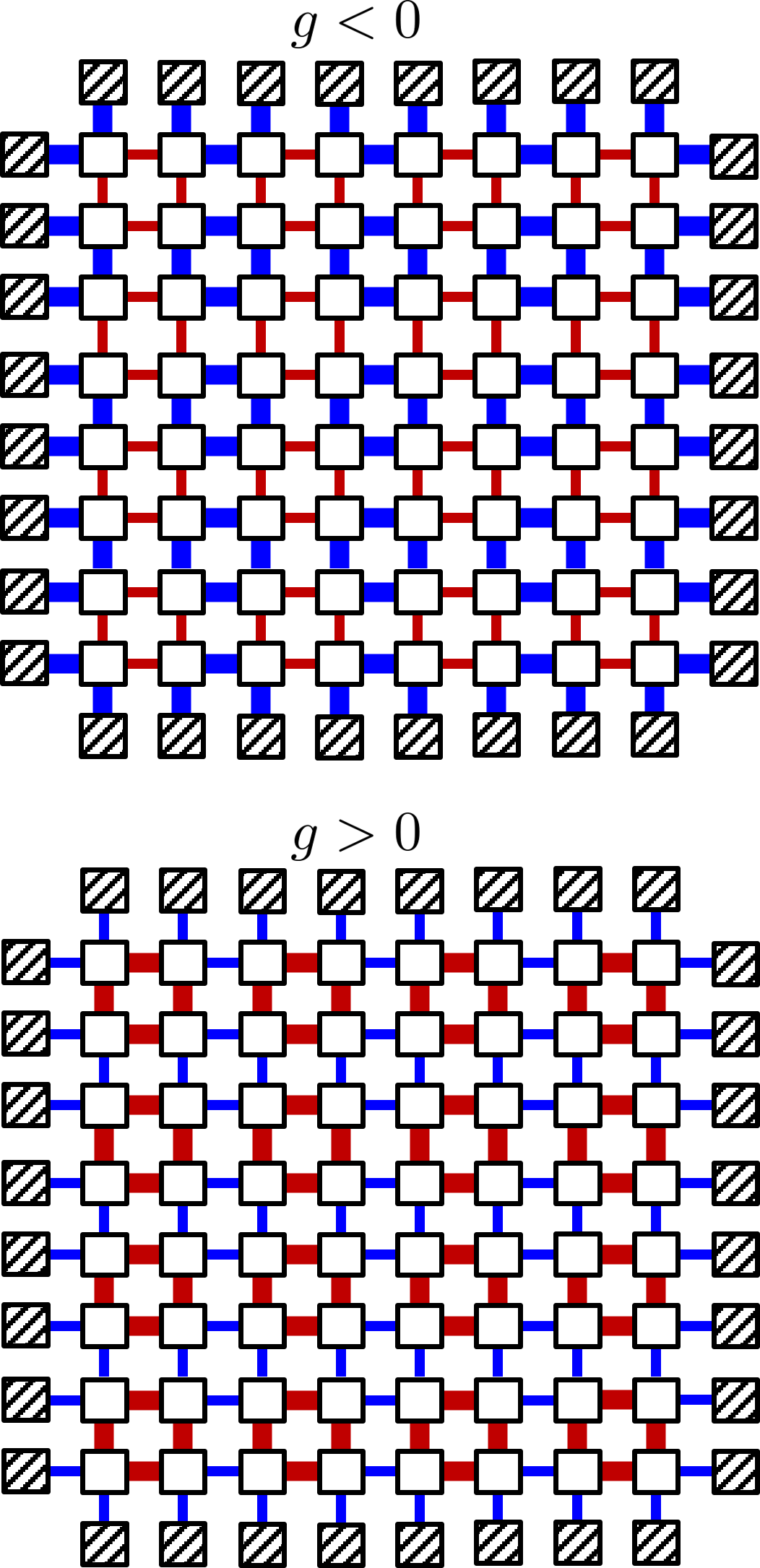}
        \label{fig:finite_geometry}
    \end{subfigure}
    \hfill
    \begin{subfigure}[t]{0.62\linewidth}
        \centering
        \caption{}
        \includegraphics[width=\linewidth]{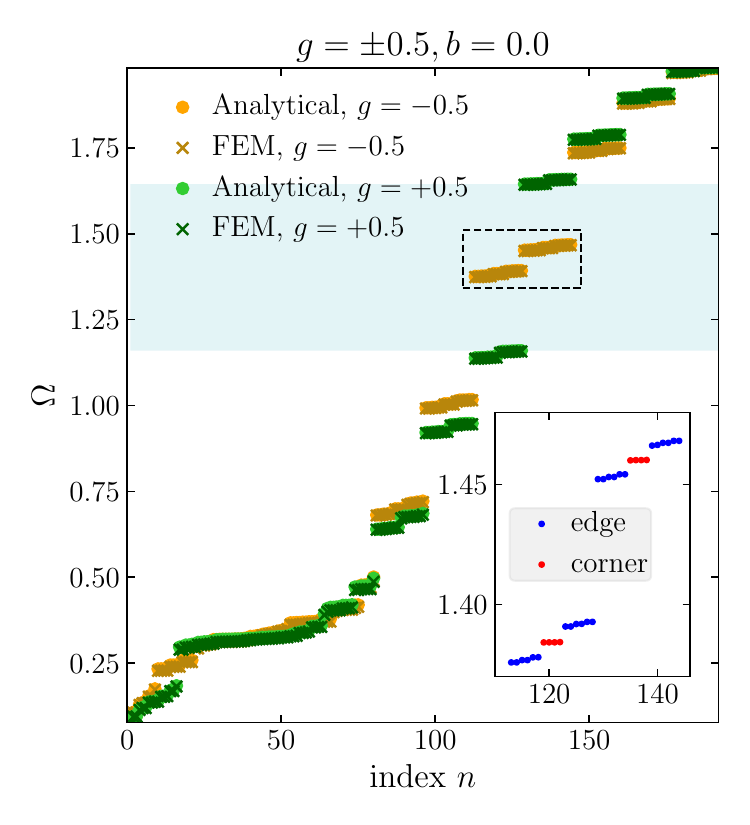}
        \label{fig:finite_eigenfrequencies}
    \end{subfigure}

    \vspace{-10mm}

    \begin{subfigure}[t]{\linewidth}
        \centering
        \caption{}
        \includegraphics[width=\linewidth]{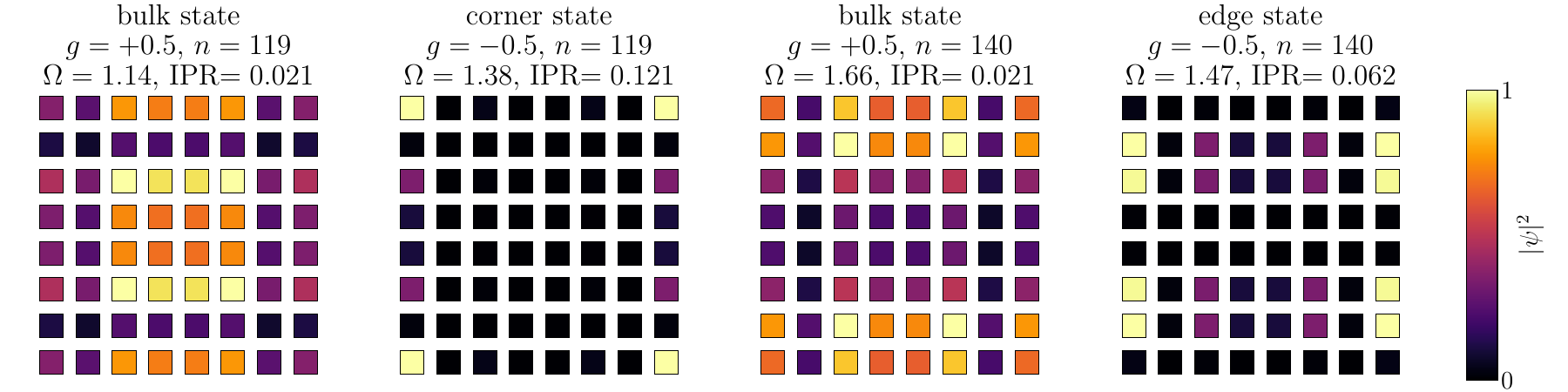}
        \label{fig:finite_maps}
    \end{subfigure}
\vspace{-5mm}
    \caption{
    (a) Geometry of the finite system with $N_x=N_y=4$.
    (b) Analytical and FEM eigenfrequencies of the finite system for the trivial and non-trivial configurations. Inset: Topological edge and corner modes inside the band gap. The dashed rectangle corresponds to the limits of the inset.
    (c) Spatial profiles of the modal weight for the topologically trivial and non-trivial configurations, showing edge- and corner-localized modes in the non-trivial sample. The corresponding mode with the same index in the trivial sample exhibits a bulk-like spatial distribution.} 
    \label{fig:main}
\end{figure}

To observe the higher order topological states, we studied a square finite system made of ${N_x\times N_y}$ UCs, with fixed boundary conditions in correspondence with the outer frame. The geometry for $N_x=N_y=4$ is shown in Fig. \ref{fig:finite_geometry} for $g>0$ and $g<0$. 

The system built from non-trivial cells displays a variety of localized modes inside the band gap (Fig. \ref{fig:finite_eigenfrequencies}). For these states, the maximum absolute displacement is localized, giving rise to topological edge and corner modes, see Fig. \ref{fig:finite_maps}. There are two sets of four degenerate corner modes, which have different frequency from the edge modes, see inset of Fig. \ref{fig:finite_eigenfrequencies}.

On the contrary, the trivial case displays an empty band gap (Fig. \ref{fig:finite_eigenfrequencies}), and therefore the modes for the same index are spatially extended throughout the bulk (Fig. \ref{fig:finite_maps}).

The theoretical model agrees with the FEM simulations, with a mean relative eigenfrequency deviation of $0.8\%$, given by
\begin{align}
\bar{\epsilon}
=
\frac{1}{N}
\sum_{n=1}^{N}
\frac{
\left|\Omega_n^{\mathrm{th}}-\Omega_n^{\mathrm{FEM}}\right|
}{
\Omega_n^{\mathrm{FEM}}
}
\times 100\%,
\end{align}
where $n$ runs across a total of $N=12N_xN_y$ eigenfrequencies.

\begin{figure}[!ht]
\centering
\includegraphics[width=\textwidth]{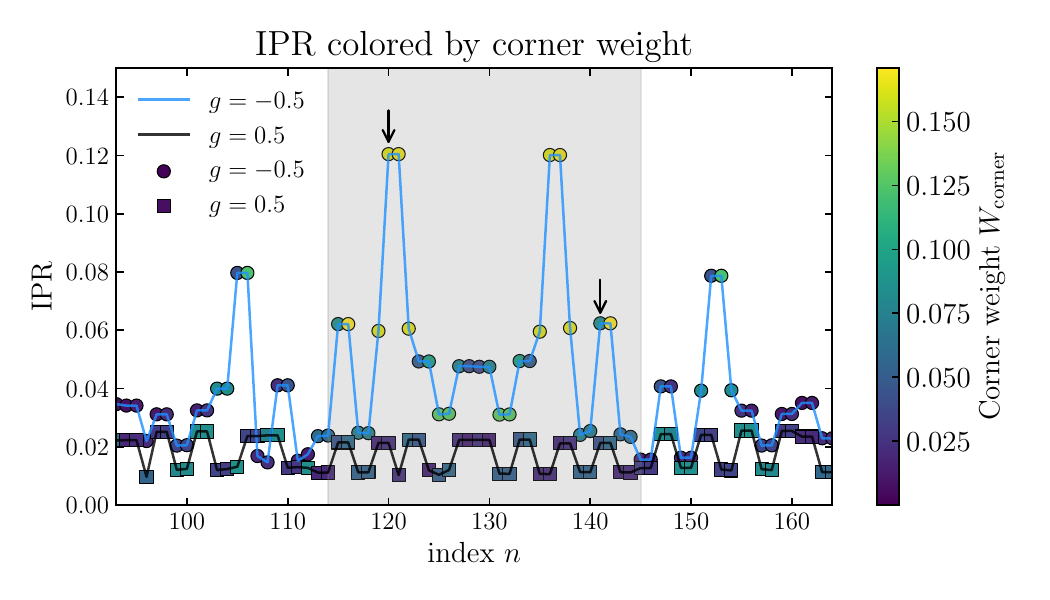}
\caption{Inverse participation ratio of the finite-system eigenmodes. The marker color denotes the normalized corner weight $W_{\mathrm{corner}}$, computed from the modal amplitude on the outermost corner squares. Modes with high IPR and high corner weight correspond to corner-localized states. The vertical arrows point to the corner and edge modes represented in Fig. \ref{fig:finite_maps}.}
\label{fig:IPR}
\end{figure}

The localization of the non-trivial modes can be quantified using the inverse participation ratio (IPR), given by
\begin{align}
    \text{IPR}=\frac{\sum_n^{N_{tot}}|\psi_n|^4}{\left(\sum_n^N|\psi_n|^2\right)^2}.
\end{align}
Modes localized on a small number of DOFs, such as topological edge and corner modes, therefore have a higher IPR than modes spread throughout the bulk.
The IPR of the modes of the finite system is shown in figure \ref{fig:IPR}, revealing a larger localization for the non-trivial modes inside the BG, with larger peaks for the corner states.
In Fig \ref{fig:IPR}, The IPR quantifies the overall localization of each eigenmode, while the color scale reports the normalized corner weight
\begin{align}
    W_{\mathrm{corner}}
    =
    \frac{\sum_{n\in\mathrm{corner}}|\psi_n|^2}
    {\sum_n|\psi_n|^2},
\end{align}
where the numerator is restricted to the degrees of freedom associated with the outermost corner squares. Modes with simultaneously large IPR and large $W_{\mathrm{corner}}$ are identified as corner-localized states.

\begin{figure}[t]
    \centering
    
    \begin{subfigure}[t]{0.42\linewidth}
        \centering
        \caption{}
        \vspace{0mm}
        \includegraphics[width=\linewidth]{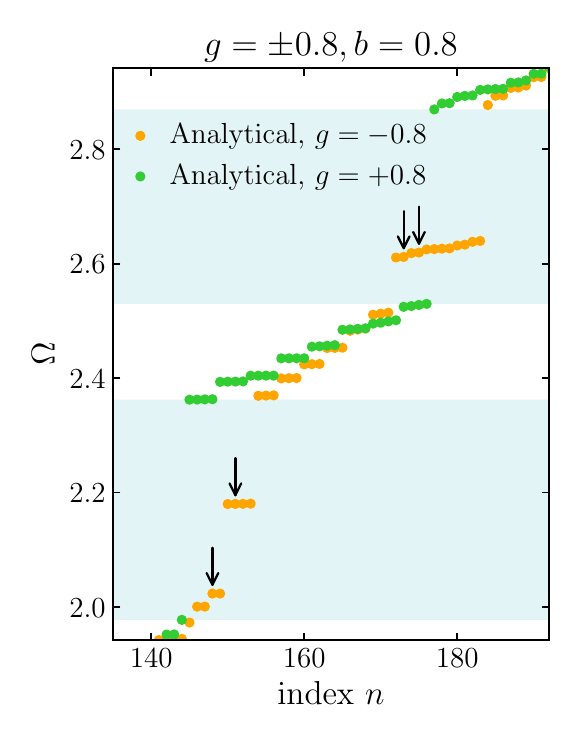}
        \label{fig:finite_dr_higher_bands}
    \end{subfigure}
    \hfill
    \hspace{-10mm}
    \begin{subfigure}[t]{0.55\linewidth}
        \centering
        \caption{}
        \includegraphics[width=\linewidth]{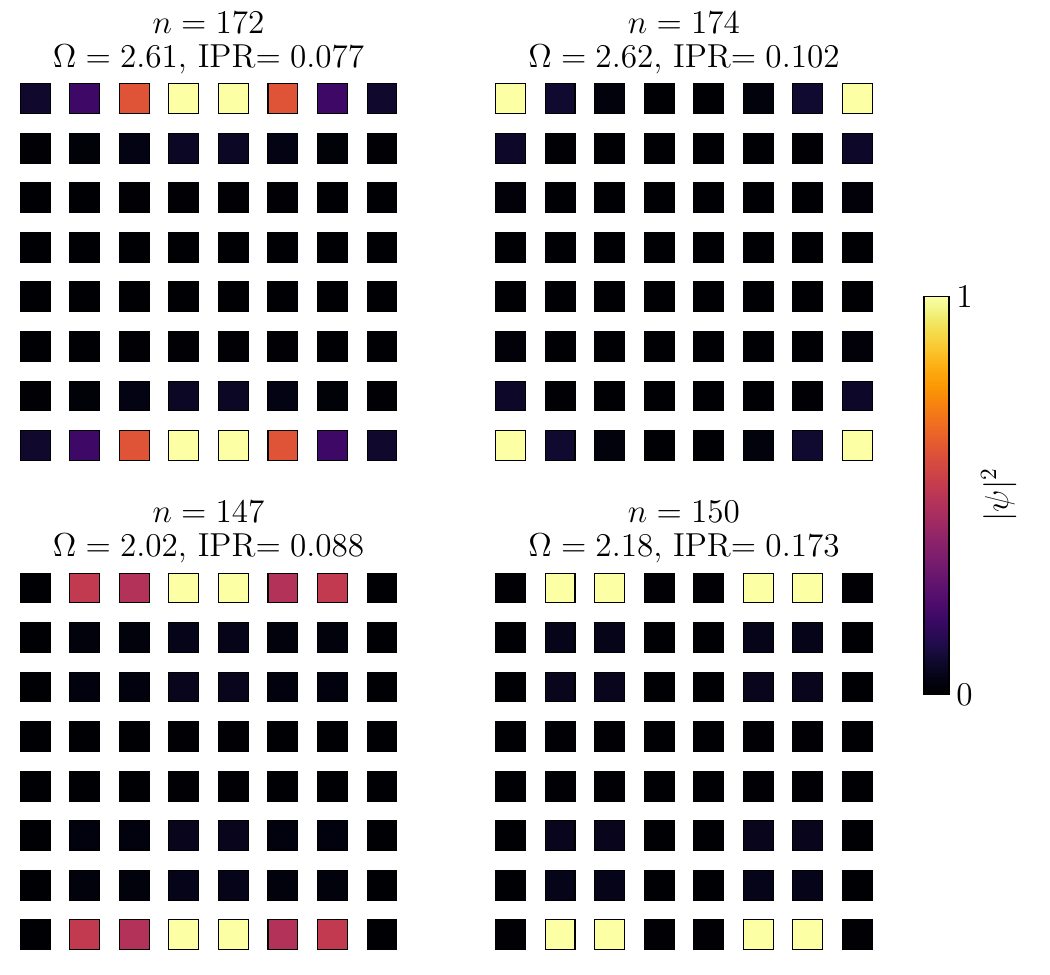}
        \label{fig:maps_higher_bands}
    \end{subfigure}
    \vspace{-5mm}
    \caption{
    (a) Analytical eigenfrequencies in the upper frequency range for the highly dimerized and off-centered configuration, $g=\pm 0.8$, $b=0.8$. The shaded regions indicate bulk gaps between bands $9$--$10$ and $11$--$12$. (b) Representative spatial profiles of in-gap modes ($g=-0.8$). The corresponding eigenfrequencies are shown by the arrows in (a).
    }
    \label{fig:higher_band_boundary_modes}
\end{figure}

For large dimerization \(|g|\) and eccentricity \(b\), several branches are shifted to higher frequencies because the off-center ligament attachments introduce additional translation--rotation coupling stiffness, as shown in Fig.~\ref{fig:dr}. These high-frequency branches open additional complete band gaps, whose cumulative bulk polarization is shown in Fig.~\ref{fig:gaps_and_polarization}. To verify the corresponding boundary response, we compute the finite spectrum for \(g=\pm0.8\) and \(b=0.8\), shown in Fig.~\ref{fig:finite_dr_higher_bands}. In agreement with the nontrivial bulk polarization for \(g<0\), in-gap modes appear in the upper gaps, whereas the corresponding \(g>0\) spectrum remains gapped. The localization of these modes appears to be gap dependent: the $9$--$10$ gap contains corner-localized modes, while the $11$--$12$ gap is dominated by edge-localized modes. Thus, although both gaps are associated with a nontrivial bulk polarization, only the lower of the two upper gaps exhibits a clear higher-order boundary response. This feature can be further exploited and will be the object of future investigations. 

\section{Experimental validation}

A simulation of an experimentally realizable finite system is performed in 3D. This includes out-of-plane effects, which are not contemplated by the analytical model. However, if the thickness is sufficiently large, the out-of-plane modes occur at frequencies much higher than the BG of interest, so that, within the relevant frequency range, we only observe the in-plane effects predicted by the analytical model.

\begin{figure}[!htbp]
    \centering

    \begin{subfigure}[t]{0.32\linewidth}
        \centering
        \caption{}
        \label{fig:setup_photo}
        \includegraphics[height=0.4\textheight]{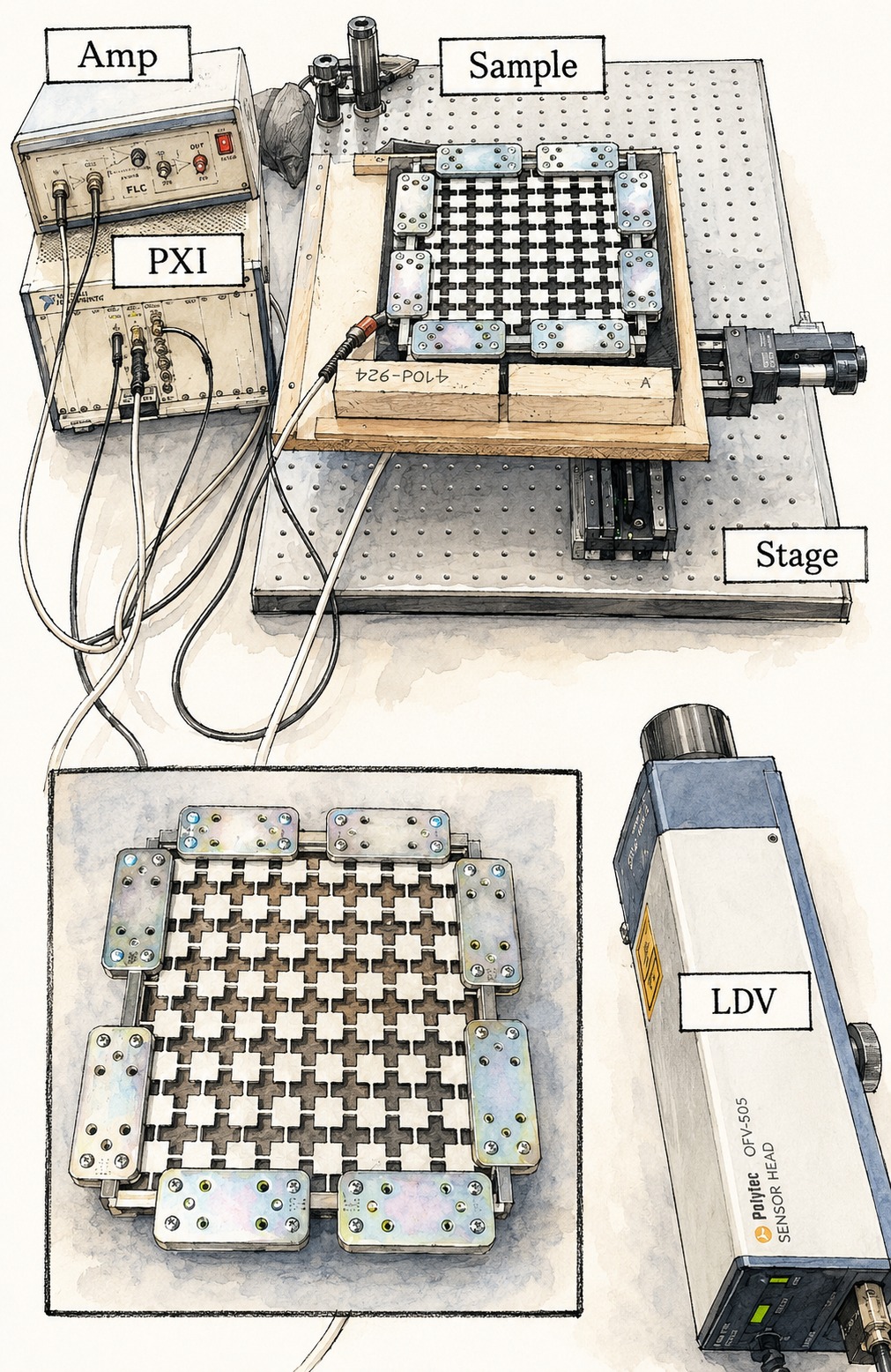}
    \end{subfigure}
    \hfill
    \begin{subfigure}[t]{0.6\linewidth}
        \centering
        \caption{}
        \label{fig:exp_spectra}
        \includegraphics[width=\linewidth]{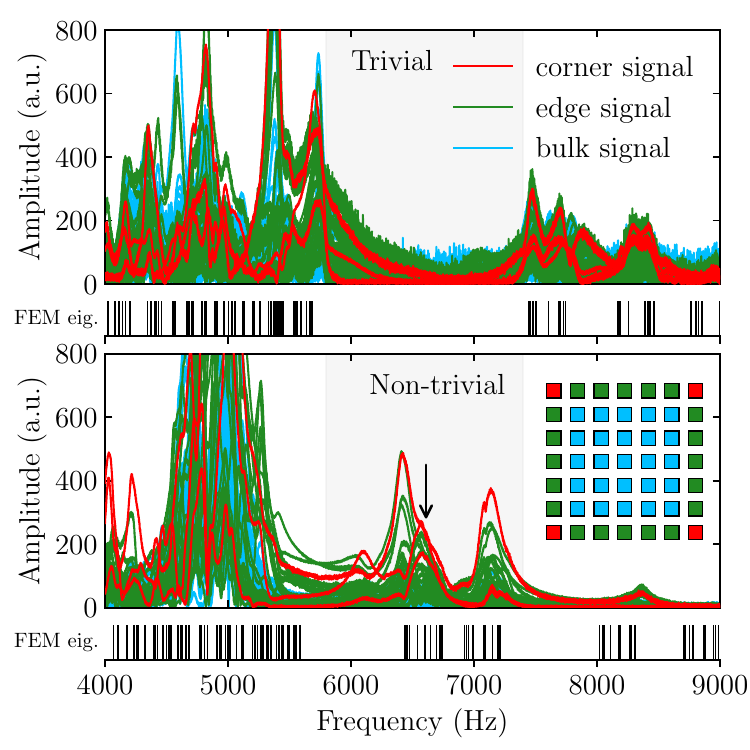}
    \end{subfigure}

    \vspace{0mm}

    \begin{subfigure}[t]{\linewidth}
        \centering
        \caption{}
        \label{fig:exp_maps}
        \includegraphics[width=\linewidth]{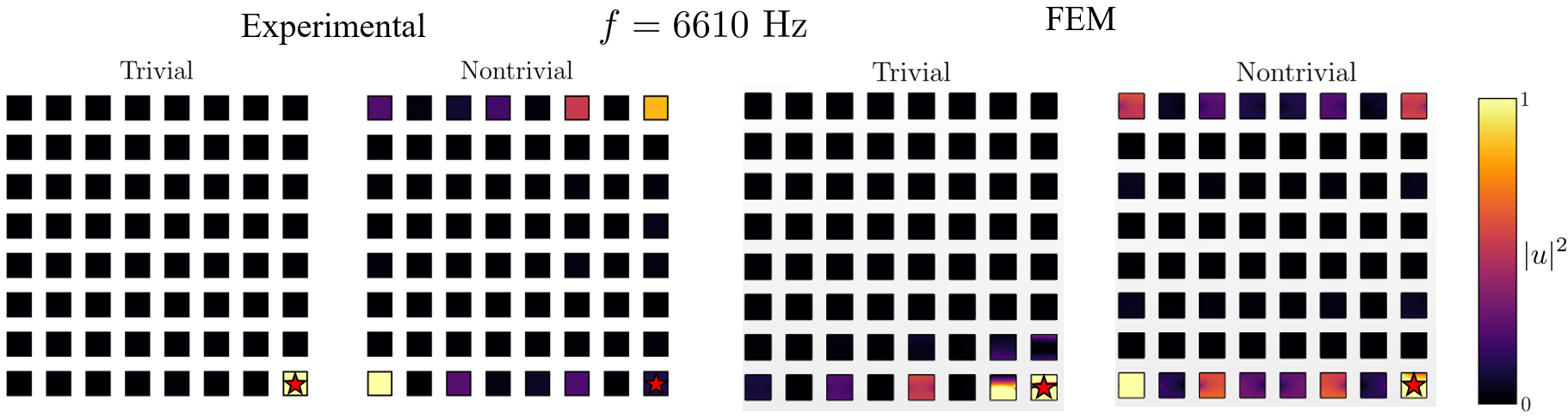}
    \end{subfigure}

    \vspace{-2mm}


 \caption{(a) Experimental setup.
(b) Measured spectra and FEM spectral densities shown below the two spectra as vertical black lines. The inset shows in red, green and blue the position of the corner, edge and bulk measured sig
(c) Experimental and numerical amplitude maps at $6610\,\mathrm{Hz}$, corresponding to the arrow in (b). The red star indicates the excited square.
The supplementary video shows the experimental amplitude map of the nontrivial sample for the entire band gap frequency range ($5800-7400$ Hz).} 
    \label{fig:complex_layout}

\end{figure}

The experimental sample consists of ${4\times4}$ UCs, surrounded by a frame (Fig. \ref{fig:setup_photo}). The sample was 3D printed in acrylonitrile 
butadiene styrene (ABS) filament. The characteristics of the sample are given in Table \ref{tab:exp}. Fixed boundary conditions were imposed by applying a clamp on the outer frame of the sample (see inset of Fig. \ref{fig:setup_photo}). The clamping was achieved using bolted perforated steel plates, separated from the sample by iron strips to avoid direct contact.

A linear sweep from $1000$ Hz to $10000$ Hz was generated, amplified and sent to a piezoelectric transducer, positioned on the side of a corner square to excite the system. The amplitudes of the in-plane vibrations were captured using a Laser Doppler Vibrometer (LDV) pointing to the side of the square. A PXI was used as interface for the signal generation and capturing. The panel was mounted on stage axis motors that translate it to the wanted $x$ and $y$ positions. A single point was measured per square and $5$ averages were performed for noise reduction (see Fig. \ref{fig:setup_photo}).

\begin{table}[]
\begin{tabular}{|l|l|}
\hline
Material                  & ABS   \\ \hline
Density $\rho$ (kg$/$m$^3$) & $1200$ \\ \hline
Young's modulus $E$ (GPa)                         & $3$                         \\ \hline
Thickness $h$ (mm)                                     & $3$                         \\ \hline
Square length $2a$ (mm)                                & $17.25$                     \\ \hline
Total sample length $L$ (mm)                                 & $240$                       \\ \hline
Beam length $l$ (mm)                                  & $10$                        \\ \hline
Average beam width $\overline{s}$ (mm)                                  & \textcolor{red}{$1.5$}                        \\ \hline
Relative width difference $g$                                 & \textcolor{red}{$0.5$}                        \\ \hline
Number of UCs $N_x\times N_y$                                    & $4\times 4$                 \\ \hline
\end{tabular}
\caption{Material and geometrical parameters of the experimental sample. The widths of the thinner and thicker beams are $1.125$ mm and $1.875$ mm, respectively.}
\label{tab:exp}
\end{table}

Two samples were fabricated, using the topologically trivial and non-trivial UCs. For the trivial sample, the amplitude spectra captured by the LDV show a sudden amplitude dip from $5800$ to ${7400 \text{ Hz}}$, corresponding to the band gap of the system. This is supported by the gap in eigenfrequencies in the same frequency range, obtained with the FEM simulation, Fig. \ref{fig:exp_spectra} (top). The experimental and FEM amplitude maps at \textcolor{red}{${6610 \text{ Hz}}$} show an evanescent wave from the excitation point, as expected for a frequency in the band gap, Fig. \ref{fig:exp_maps}. The spectra of the non-trivial sample exhibit distinct peaks in the band gap region. The squares located at the corners (red) and along the edges (green) of the sample have the highest amplitude, Fig. \ref{fig:exp_spectra}. These peaks are supported by an increased density of eigenstates of the FEM simulation, represented by the vertical black lines in Fig. \ref{fig:exp_spectra}, shown below the two spectra. The same excitation at \textcolor{red}{${6610 \text{ Hz}}$} in the non-trivial sample produces amplitude maxima on squares located on opposite sides from the excitation point, highlighting an edge-mediated driven response, that is focused on the corners and caused by the presence of a non-trivial topological corner state at that frequency. The experimental and FEM \textcolor{red}{driven} responses contain respectively 25.4\% and 29.5\% of the total amplitude at the corners opposite from the excitation point, compared with 33.2\% of a corresponding analytical eigenmode.  This indicates that the driven responses share spatial localization features with the analytical eigenmode.
Appendix \ref{sec:app_decay} provides a more detailed comparison of the analytical eigenmode and the measured and simulated driven responses at ${6610 \text{ Hz}}$. 

Several edge and corner modes 
occur
at nearby frequencies. The experimentally measured response 
reflects their combined contributions, with relative weights determined by the applied excitation and damping.  The supplementary video shows how the spatial distribution of the driven vibration amplitude varies with excitation frequency across the band gap, from ${5800 \text{ Hz}}$ to ${7400 \text{ Hz}}$.


\newsavebox{\rightblock}
\FloatBarrier

\section{Conclusions}

We have shown that a continuous elastic phononic crystal with ligament-connected rigid units can be reduced to a model described by a compact Hermitian dynamical matrix that retains the essential mechanics needed to predict higher-order topology. Unlike ideal mass–spring SSH models, the reduction includes rotational degrees of freedom, bending stiffness, and ligament eccentricity, allowing purely geometric end elastic parameters to control both the spectral gap and the topological phase. The model predicts the bulk polarization, the transition between trivial and non-trivial phases, and the emergence of edge and corner modes in finite structures.

The localization is quantified by an exponential fit that matches the \textcolor{red}{expected} decay length of the SSH model. The agreement with FEM and experimental measurements on 3D-printed samples demonstrates that the reduction is not only a conceptual description but a practical design tool for manufacturable topological elastic metamaterials.

In the context of the existing literature, our work introduces a mechanically explicit beam-lattice model that retains not only in-plane translations but also rigid-body rotations, ligament bending, and attachment eccentricity. Moreover, the multiband character of our beam-lattice reduction reveals a topological structure that cannot be captured by either a scalar or a purely translational vectorial SSH description. The 8--9
gap is the clearest example: its cumulative bulk polarization is
trivial, yet it supports four ribbon edge states and eight corner
modes. We showed that this is resolved at the level of the mirror
sectors $\sigma_{tb}$ and $\sigma_{lr}$, each of which undergoes an
independent SSH-type inversion and carries a non-trivial
sector-resolved Zak phase that cancels only in the cumulative
invariant. Boundary and corner localization is thus governed by this
sector-resolved topology, not by the net polarization. Similarly, the
ligament eccentricity $b$, by enhancing translation--rotation
coupling, can reverse which gaps (7--8, 9--10, 11--12) are non-trivial
as $g$ changes sign, showing that the rotational degrees of freedom
retained by the model actively shape the topology, rather than being a
minor correction. This indicates that higher-order topological classification in
continuum-derived lattices should therefore generally be performed at
the level of symmetry sectors, particularly whenever rotational or eccentricity-related coupling is non-negligible. The off-centred configurations investigated here demonstrate the ability of the proposed approach to capture eccentricity-dependent changes in bulk topology and boundary localization. However, the primary aim of this work is to establish and validate the modelling approach, rather than to provide an exhaustive characterization of the  effects that are accessible for $b\neq 0$. A systematic investigation of these configurations and their possible connections with topological behaviour is reserved for a dedicated study. 

The considered system, despite its simplicity, can act as a basic building block for a wide range of different metamaterial systems. This framework can therefore be used for rapid phase-space exploration, inverse design of localized vibration modes, and future extensions to three-dimensional or tunable phononic architectures.


\begin{acknowledgments}

\end{acknowledgments}
G.C. and M.B. acknowledge the support of the Fondazione di Sardegna through the project FdS 2023 "Smart Materials and Morphing Systems for efficient energy harvesting based on fluid-structure interactions" (F23C25000280007).

\newpage
\section*{References}
\bibliography{bibliography}

\newpage
\appendix

\section{Derivation of governing equations}\label{app:Forces-and-Moments}
\label{AppA}
\setcounter{equation}{0}
\renewcommand{\theequation}{A.\arabic{equation}}
\setcounter{figure}{0}
\renewcommand{\thefigure}{A.\arabic{figure}}
 
Here, we show how to determine the equations of motion for the rigid squares of mass $m$ and moment of inertia $I$. In particular, as an example, we focus our attention on the square denoted as A in Fig. \ref{fig:uc}, located within the unit cell labelled as ($p,q$).

The linear and angular momentum balance equations for the rigid square A, obtained by equating the inertial forces and moments to the elastic contributions provided by the elastic beams, are given by
\begin{subequations}\label{EqsMotion}
\begin{align}
\nonumber
& m \ddot{u}^{\textup{A}}_{p,q} = \underbrace{\frac{E A_2}{l} \left( u^{\textup{B}}_{p,q}-u^{\textup{A}}_{p,q} \right)}_{F_1} + \frac{E A_1}{l} \left( u^{\textup{B}}_{p-1,q}-u^{\textup{A}}_{p,q} \right) + \frac{12 E J_6}{l^3} \left( u^{\textup{C}}_{p,q}-u^{\textup{A}}_{p,q} \right) \\
\nonumber
& + \frac{12 E J_5}{l^3} \left( u^{\textup{C}}_{p,q-1}-u^{\textup{A}}_{p,q} \right) + \left( \frac{6 E J_6}{l^2} - \frac{6 E J_5}{l^2} + \frac{12 E J_6}{l^3}a - \frac{12 E J_5}{l^3}a + \frac{E A_1}{l}b \underbrace{- \frac{E A_2}{l}b}_{F_8} \right) \theta^{\textup{A}}_{p,q} \\
& + \underbrace{\frac{E A_2}{l}b\, \theta^{\textup{B}}_{p,q}}_{F_5} - \frac{E A_1}{l}b\, \theta^{\textup{B}}_{p-1,q} + \frac{6 E J_6}{l^2} \left( 1+\frac{2 a}{l} \right) \theta^{\textup{C}}_{p,q} - \frac{6 E J_5}{l^2} \left( 1+\frac{2 a}{l} \right) \theta^{\textup{C}}_{p,q-1} \, , \\
\nonumber
& m \ddot{v}^{\textup{A}}_{p,q} = \frac{E A_6}{l} \left( v^{\textup{C}}_{p,q}-v^{\textup{A}}_{p,q} \right) + \frac{E A_5}{l} \left( v^{\textup{C}}_{p,q-1}-v^{\textup{A}}_{p,q} \right) + \underbrace{\frac{12 E J_2}{l^3} \left( v^{\textup{B}}_{p,q}-v^{\textup{A}}_{p,q} \right)}_{F_2} \\
\nonumber
& + \frac{12 E J_1}{l^3} \left( v^{\textup{B}}_{p-1,q}-v^{\textup{A}}_{p,q} \right) + \left( \frac{6 E J_1}{l^2} \underbrace{- \frac{6 E J_2}{l^2}}_{F_6} + \frac{12 E J_1}{l^3}a \underbrace{- \frac{12 E J_2}{l^3}a}_{F_7} + \frac{E A_5}{l}b - \frac{E A_6}{l}b \right) \theta^{\textup{A}}_{p,q} \\
& \underbrace{- \frac{6 E J_2}{l^2} \left( 1+\frac{2 a}{l} \right) \theta^{\textup{B}}_{p,q}}_{F_3+F_4} + \frac{6 E J_1}{l^2} \left( 1+\frac{2 a}{l} \right) \theta^{\textup{B}}_{p-1,q} + \frac{E A_6}{l}b\, \theta^{\textup{C}}_{p,q} - \frac{E A_5}{l}b\, \theta^{\textup{C}}_{p,q-1} \, , \\
\nonumber
&I \ddot{\theta}^{\textup{A}}_{p,q} = - \left( \frac{4 E J_1}{l}+\underbrace{\frac{4 E J_2}{l}}_{M_6^{\textup{I}}}+\frac{4 E J_5}{l}+\frac{4 E J_6}{l} + \frac{12 E J_1}{l^2}a+\underbrace{\frac{12 E J_2}{l^2}a}_{M_6^{\textup{II}}+M_7^{\textup{I}}}+\frac{12 E J_5}{l^2}a+\frac{12 E J_6}{l^2}a \right. \\
\nonumber
& \left. \frac{12 E J_1}{l^3}a^2+\underbrace{\frac{12 E J_2}{l^3}a^2}_{M_7^{\textup{II}}}+\frac{12 E J_5}{l^3}a^2+\frac{12 E J_6}{l^3}a^2 + \frac{E A_1}{l}b^2+\underbrace{\frac{E A_2}{l}b^2}_{M_8}+\frac{E A_5}{l}b^2+\frac{E A_6}{l}b^2 \right) \theta^{\textup{A}}_{p,q} \\
\nonumber
& \underbrace{-\left[ \frac{2 E J_2}{l} \left( 1 + \frac{6 a}{l} + \frac{6 a^2}{l^2} \right) - \frac{E A_2}{l}b^2 \right] \theta^{\textup{B}}_{p,q}}_{M_3+M_4+M_5} -\left[ \frac{2 E J_1}{l} \left( 1 + \frac{6 a}{l} + \frac{6 a^2}{l^2} \right) - \frac{E A_1}{l}b^2 \right] \theta^{\textup{B}}_{p-1,q} \\
\nonumber
& -\left[ \frac{2 E J_6}{l} \left( 1 + \frac{6 a}{l} + \frac{6 a^2}{l^2} \right) - \frac{E A_6}{l}b^2 \right] \theta^{\textup{C}}_{p,q} -\left[ \frac{2 E J_5}{l} \left( 1 + \frac{6 a}{l} + \frac{6 a^2}{l^2} \right) - \frac{E A_5}{l}b^2 \right] \theta^{\textup{C}}_{p,q-1} \\
\nonumber
& \underbrace{+\frac{E A_2}{l}b \left( u^{\textup{B}}_{p,q}-u^{\textup{A}}_{p,q} \right)}_{M_1} -\frac{E A_1}{l}b \left( u^{\textup{B}}_{p-1,q}-u^{\textup{A}}_{p,q} \right) +\frac{E A_6}{l}b \left( v^{\textup{C}}_{p,q}-v^{\textup{A}}_{p,q} \right) -\frac{E A_5}{l}b \left( v^{\textup{C}}_{p,q-1}-v^{\textup{A}}_{p,q} \right) \\
\nonumber
& -\frac{6 E J_6}{l^2} \left( 1 + \frac{2 a}{l} \right) \left( u^{\textup{C}}_{p,q}-u^{\textup{A}}_{p,q} \right) +\frac{6 E J_5}{l^2} \left( 1 + \frac{2 a}{l} \right) \left( u^{\textup{C}}_{p,q-1}-u^{\textup{A}}_{p,q} \right) \\
& \underbrace{+\frac{6 E J_2}{l^2} \left( 1 + \frac{2 a}{l} \right) \left( v^{\textup{B}}_{p,q}-v^{\textup{A}}_{p,q} \right)}_{M_2} -\frac{6 E J_1}{l^2} \left( 1 + \frac{2 a}{l} \right) \left( v^{\textup{B}}_{p-1,q}-v^{\textup{A}}_{p,q} \right) \, , \label{BlHam2}
\end{align}
\end{subequations}
where $E$ is the Young's modulus, $l$ is the length, $A_i$ and $J_i$ ($i=1,2,5,6$ in this case) are the cross-sectional area and second moment of area of the beams connecting square A with the neighbouring squares. In addition, $a$ denotes the half-length of the square and $b$ is the eccentricity (see Fig. \ref{fig:uc}).

\begin{figure}[!ht]
\centering
\includegraphics[width=0.8\textwidth]{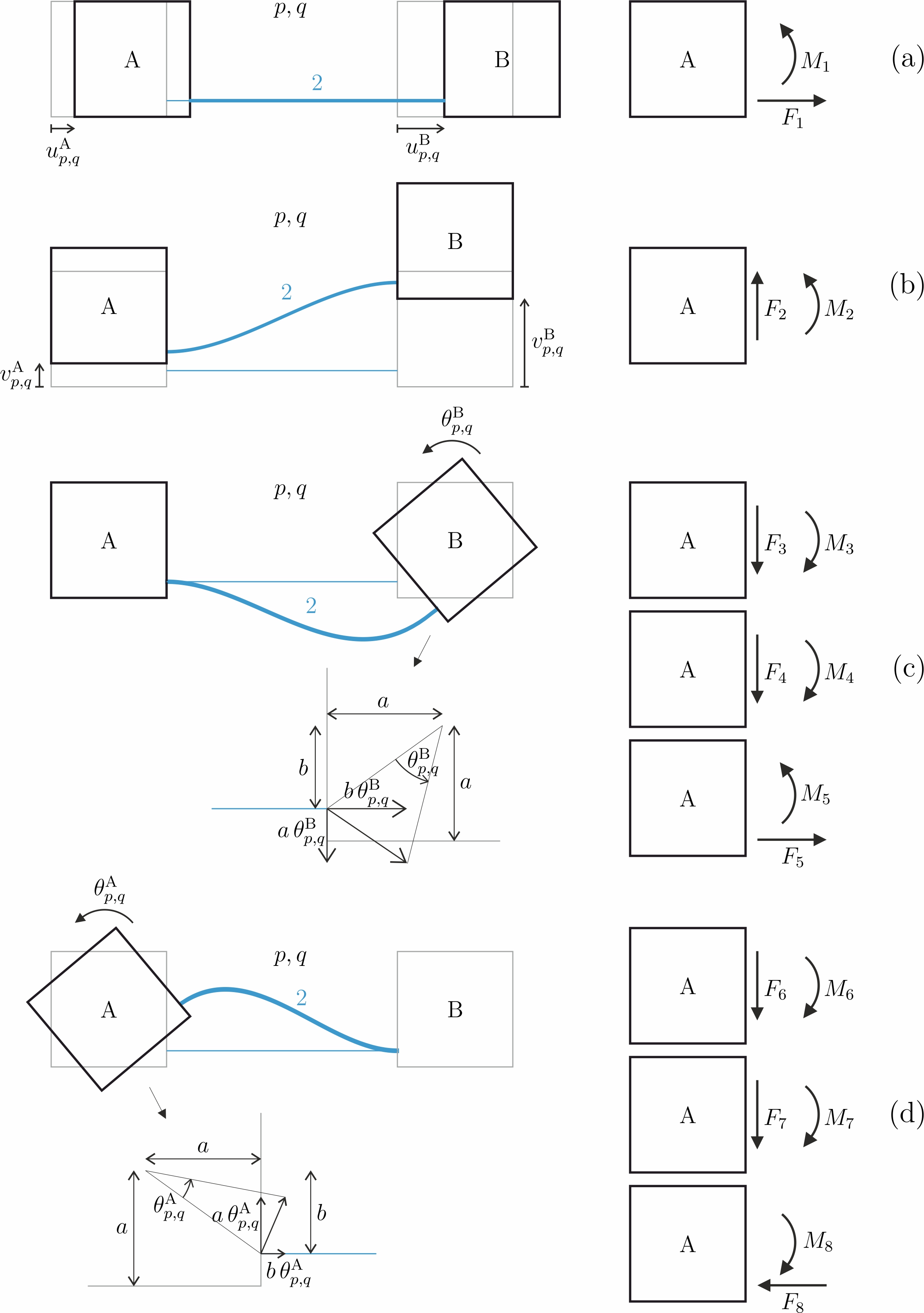}
\caption{Forces and moments on square A due to the relative motion with square B, which is connected to square A by the elastic beam labeled 2: relative displacement in the $x$ direction (a), relative displacement in the $y$ direction (b), rotation of square B (c) and rotation of square A (d).}
\label{fig:FreeBodyDiagrams}
\end{figure}

\begin{figure}[!ht]
\centering
\includegraphics[width=0.9\textwidth]{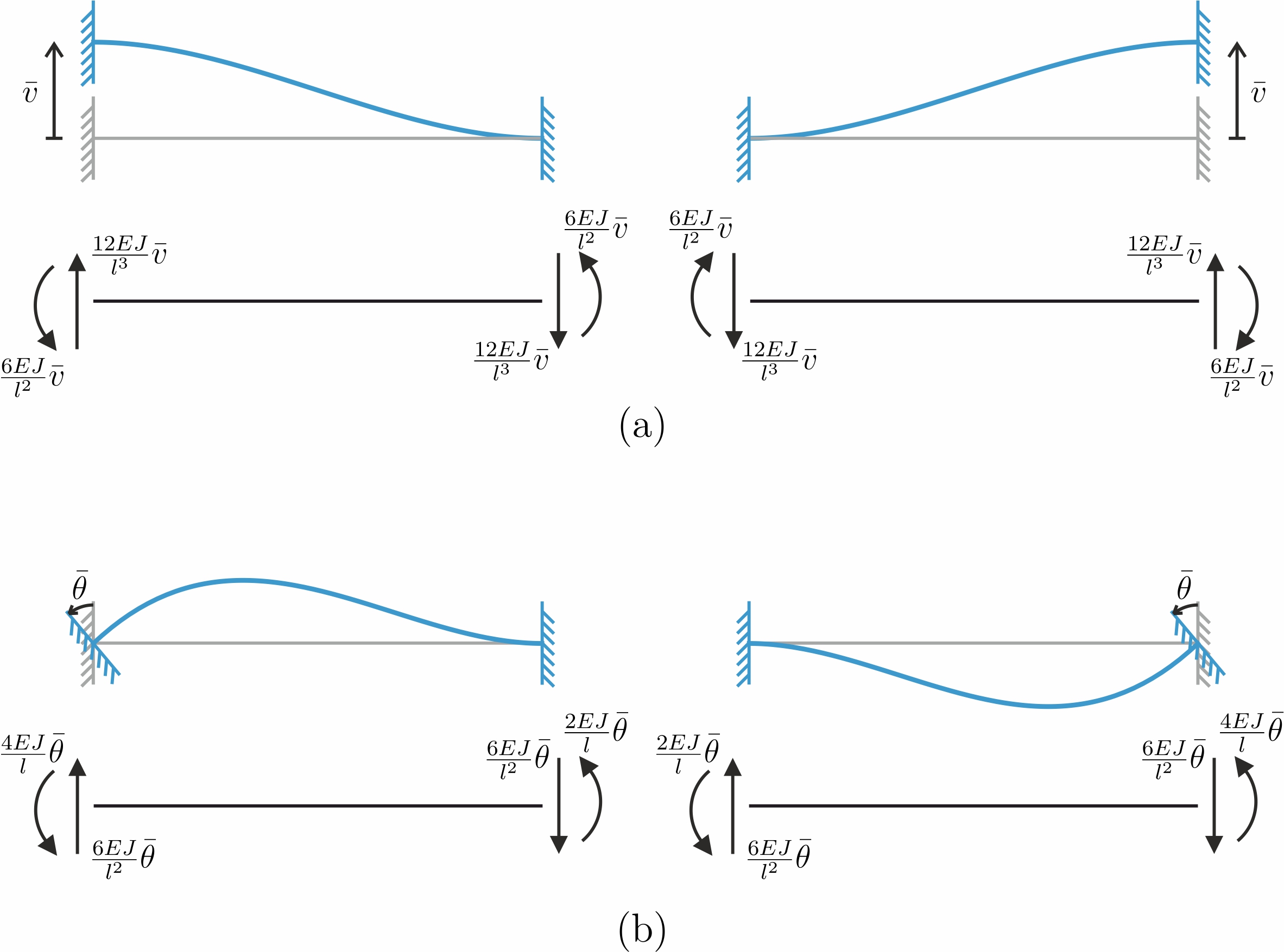}
\caption{Forces and moments at the edges of a clamped-clamped beam, due to an imposed transverse displacement $\bar{v}$ (a) and an imposed rotation $\bar{\theta}$ (b) at either end of the structure (left/right panels).}
\label{fig:Forces_Moments}
\end{figure}

In the following, we derive the expressions of the elastic terms, underbraced in Eqs. (\ref{EqsMotion}). They are obtained by considering the relative motion between squares A and B, as shown in Fig. \ref{fig:FreeBodyDiagrams}. To this end, we also exploit the free-body diagrams of the statically indeterminate structure, shown in Fig. \ref{fig:Forces_Moments}. The remaining elastic contributions in Eqs. (\ref{EqsMotion}) can be determined following a similar approach.
The model is derived under linear, elastic, isotropic, and small-displacement assumptions and treats the square inclusions as rigid bodies connected by Euler–Bernoulli ligaments. Therefore, shear deformation, rotary inertia of the ligaments, material damping, printing-induced anisotropy, and out-of-plane motion are not included in the analytical dynamical matrix. The comparison with FEM indicates that these effects are small in the frequency range of the target band gap, but they are expected to become relevant for thinner samples, shorter ligaments, or higher bands. 

When squares A and B undergo a relative displacement along the $x$ direction, square A is subjected to a force $F_1$ due to the elongation of beam 2 and to the resulting moment arising from the force eccentricity:
\begin{equation}
F_1 = \frac{E A_2}{l} \left( u^{\textup{B}}_{p,q}-u^{\textup{A}}_{p,q} \right) \,, \qquad M_1 = F_1 \, b \, .
\end{equation}
A relative displacement in the $y$ direction generates a force and a moment acting on square A, obtained from the free-body diagram in Fig. \ref{fig:Forces_Moments}(a) and accounting for the force offset:
\begin{equation}\label{eq:Forces_moments_F2_to_F5}
F_2 = \frac{12 E J_2}{l^3} \left( v^{\textup{B}}_{p,q}-v^{\textup{A}}_{p,q} \right) \,, \qquad M_2 = \frac{6 E J_2}{l^2} \left( v^{\textup{B}}_{p,q}-v^{\textup{A}}_{p,q} \right) + F_2 \, a \, .
\end{equation}
When square B undergoes a rotation, forces and moments act on square A, arising not only from the rotation itself but also from the displacement of the point where beam 2 is attached to square B (see Fig. \ref{fig:Forces_Moments}):
\begin{equation}
\begin{aligned}
&F_3 = -\frac{6 E J_2}{l^2} \theta^{\textup{B}}_{p,q} \,, \qquad M_3 = -\frac{2 E J_2}{l} \theta^{\textup{B}}_{p,q} - F_3 \, a \, , \\
&F_4 = -\frac{12 E J_2}{l^3}a\, \theta^{\textup{B}}_{p,q} \,, \qquad M_4 = -\frac{6 E J_2}{l^2}a\, \theta^{\textup{B}}_{p,q} - F_4 \, a \, , \\
&F_5 = \frac{E A_2}{l}b\, \theta^{\textup{B}}_{p,q} \,, \qquad M_5 = F_5 \, b \, .
\end{aligned}
\end{equation}
Finally, the rotation of square A generates forces and moments on the same square, which can be determined following a similar approach as above:
\begin{equation}
\begin{aligned}
&F_6 = -\frac{6 E J_2}{l^2} \theta^{\textup{A}}_{p,q} \,, \qquad M_6 = -\frac{4 E J_2}{l} \theta^{\textup{A}}_{p,q} - F_6 \, a = M_6^{\textup{I}} + M_6^{\textup{II}} \, , \\
&F_7 = -\frac{12 E J_2}{l^3}a\, \theta^{\textup{A}}_{p,q} \,, \qquad M_7 = -\frac{6 E J_2}{l^2}a\, \theta^{\textup{A}}_{p,q} - F_7 \, a = M_7^{\textup{I}} + M_7^{\textup{II}} \, , \\
&F_8 = -\frac{E A_2}{l}b\, \theta^{\textup{A}}_{p,q} \,, \qquad M_8 = F_8 \, b \, .
\end{aligned}
\end{equation}

\section{Separation of the bands in odd-even sectors}\label{sec:separation_odd_even}

To identify the topological structure of the multiband spectrum, we analyze the mirror symmetries of the dynamical matrix $H(\mathbf{k})$.
For $b=0$, the unit cell has $C_{4v}$ symmetry and is invariant
under top-bottom and left-right mirror reflections. We denote the
corresponding operators by \(S_{tb}\) and \(S_{lr}\).
These operators act on the mechanical DOFs as
\begin{align}
    S_{tb}: \begin{cases}
        u_A\leftrightarrow u_C,\quad v_A\leftrightarrow -v_C,\quad \theta_A\leftrightarrow-\theta_C,\\
        u_B\leftrightarrow u_D,\quad v_B\leftrightarrow -v_D,\quad \theta_B\leftrightarrow-\theta_D. \\
    \end{cases}
    \\[10 pt]
    S_{lr}: \begin{cases}
    u_A\leftrightarrow -u_B,\quad v_A\leftrightarrow v_B, \quad \theta_A\leftrightarrow-\theta_B, \\
    u_C\leftrightarrow -u_D,\quad v_C\leftrightarrow v_D, \quad \theta_C\leftrightarrow-\theta_D. \\
    \end{cases}
\end{align}
Both operators satisfy \(S_{tb}^2=S_{lr}^2=I\), and therefore have eigenvalues $\sigma_{tb}=\pm 1$ and $\sigma_{lr}=\pm 1$. Modes with positive eigenvalue are mirror even while modes with negative eigenvalue are mirror odd.

The mirror symmetries impose the following relations on the dynamical matrix:
\begin{align}
    S_{tb}H(k_x,k_y)=H(k_x,-k_y)S_{tb},\\ S_{lr}H(k_x,k_y)=H(-k_x,k_y)S_{lr}.
\end{align}

Thus, on the mirror-invariant lines \(k_y=0,\pi/d\),
\([H,S_{tb}]=0\), while on the mirror-invariant lines \(k_x=0,\pi/d\),
\([H,S_{lr}]=0\). On these lines, the eigenmodes of the dynamical matrix can be chosen to be also eigenmodes of the corresponding mirror operators, hence they are necessarily even or odd. 
Equivalently, the Hilbert space decomposes into independent even and
odd sectors:
\begin{align}
\mathcal{H}=\mathcal{H}_{+}\oplus\mathcal{H}_{-},
\end{align}
and the dynamical matrix block diagonalizes as
\begin{align}
H=H_{+}\oplus H_{-}.
\end{align}
Consequently, bands belonging to opposite mirror sectors cannot
hybridize, and crossings between them do not affect the Zak phase
computed within a gapped sector.

Indeed, each sector has a gapped band structure with a non-trivial Zak phase. This non-trivial topology justifies the presence of four ribbon edge states. For each of the two edges, there is an odd-even pair of localized states.

\section{Ribbon modes for $b\neq 0$}\label{sec:app:ribbon}

\setcounter{equation}{0}
\renewcommand{\theequation}{C.\arabic{equation}}
\setcounter{figure}{0}
\renewcommand{\thefigure}{C.\arabic{figure}}

\begin{figure}
\centering
    
    \vfill
    \vspace{-10mm}
    
    \begin{subfigure}[b]{0.49\textwidth}
    \centering
    \caption{\label{fig:image2}}
    \label{fig:ribbon_b08_changing_g}
    \vspace{-2mm}
    \includegraphics[width=\textwidth]{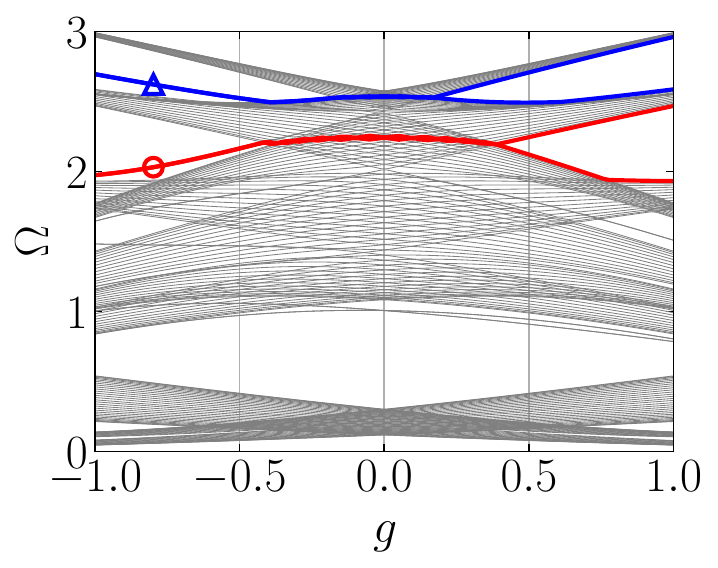}
    \end{subfigure}
    \hfill
    \begin{subfigure}[b]{0.49\textwidth}
    \centering
    \caption{\label{fig:image3}}
    \label{fig:ribbon_b08_states}
    \includegraphics[width=1\textwidth]{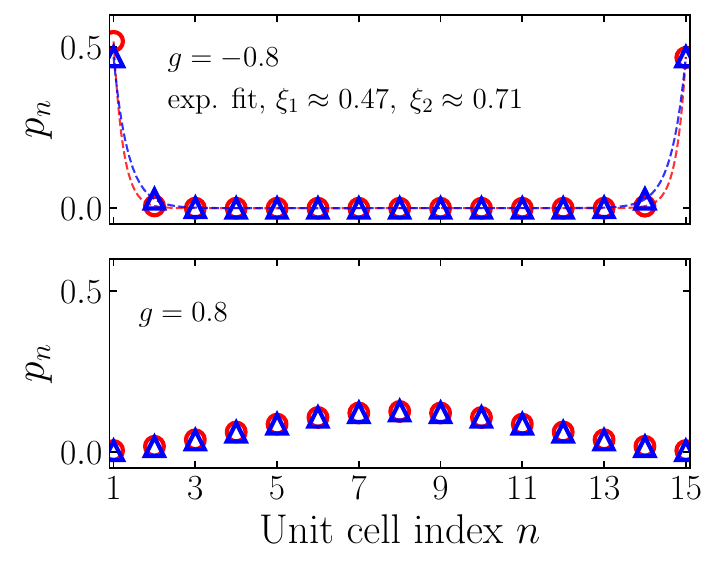}
    \end{subfigure}
    
    
    
\caption{
(\subref{fig:image2}) Ribbon eigenfrequencies for varying $g$ for $N_x=15$ unit cells. The red and blue lines correspond to pair of degenerate states which are in-gap for $g<0$ and belong to the bulk for $g>0$. (\subref{fig:image3}) (Top) Spatial profile\textcolor{red}{s} and exponential fit\textcolor{red}{s} of the ribbon edge states for $g=-0.8$, (\textcolor{red}{b}ottom) spatial profile\textcolor{red}{s} of the bulk states with the same index at $g=+0.8$. 
}
\label{fig:ribbon_b08}
\end{figure}

The analytical results of the ribbon off-centered structure with $b=0.8$ are shown in Fig. \ref{fig:ribbon_b08}. The spectrum for changing $g$ in Fig. \ref{fig:ribbon_b08_changing_g} displays two degenerate pairs of in-gap states for high negative dimerization $\lesssim -0.4$  (red and blue lines), which join the bulk for positive $g$. The markers correspond to the states whose profile\textcolor{red}{s are} displayed in Fig. \ref{fig:ribbon_b08_states}, with a clear distinction between in-gap localized edge states for $g<0$ and bulk modes for $g>0$. 

\section{Intensity decay of corner states}\label{sec:app_decay}

\setcounter{equation}{0}
\renewcommand{\theequation}{D.\arabic{equation}}
\setcounter{figure}{0}
\renewcommand{\thefigure}{D.\arabic{figure}}

\begin{figure}[ht!]
    \centering
    \includegraphics[width=\linewidth]{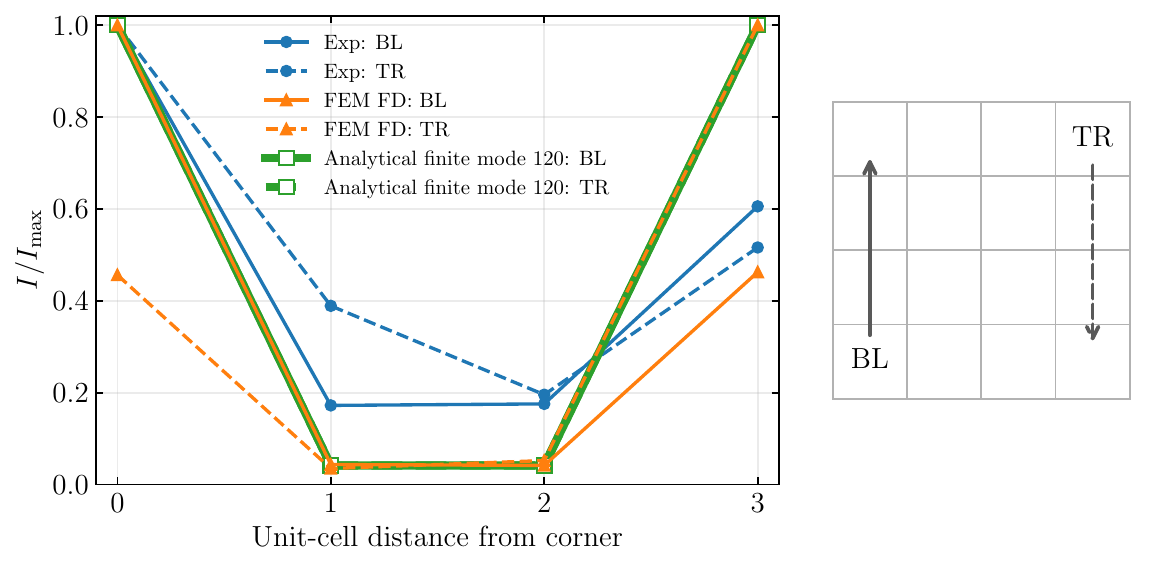}
    \caption{Normalized unit-cell resolved intensity of the driven response and analytical finite and ribbon eigenmodes for a corner driven response at 6610 Hz from the top right (TR) and bottom left (BL) corners. Here, $I$ is the sum of the absolute squared displacement over a unit cell, and $I_\mathrm{max}$ is the maximum intensity along each path.}
    \label{fig:UC-resolved-intensity-comparison}
\end{figure}

Figure \ref{fig:UC-resolved-intensity-comparison} compares the unit cell-resolved intensity of the driven corner state at $6610$ Hz, with that of the corresponding analytically calculated eigenmode. The driven \textcolor{red}{response} and eigenmode results show a comparable decay in intensity over the first and second unit cells away from the corner. 
The analytical eigenmodes have maximum intensity at their four corners, which explains why the amplitude rises back to $1$ at the end of the path. This is not the case for the driven reponses,
where opposite corners  show a lower amplitude.
The small number of unit cells makes it difficult to reliably calculate a decay length: as intensity decreases, it increases near another corner. Material damping (the sample is manufactured using a polymer) may broaden the resonances, allowing nearby eigemodes with different spatial characteristics to contribute to the driven response and modify its 
spatial profile.

\end{document}